\documentclass[pdftex,twocolumn,epjc3]{svjour3}          

\RequirePackage[T1]{fontenc}

\RequirePackage{graphicx}
\RequirePackage{mathptmx}   
\RequirePackage{flushend}
\RequirePackage[numbers,sort&compress]{natbib}
\RequirePackage[colorlinks,citecolor=blue,urlcolor=blue,linkcolor=blue]{hyperref}

\usepackage{graphicx}
\usepackage{dcolumn}
\usepackage{bm}
\usepackage{amssymb}
\usepackage{enumerate}
\usepackage{lineno}
\usepackage{textalpha} 
\usepackage{doi} 

\newcommand\ee{e^+e^-}

\newcommand\aee{A' \to e^+e^-}
\newcommand\xdecay{X17 \rightarrow e^+  e^-}

\newcommand\pair{e^+e^-}

\newcommand\emu{e^- Z \to e^- Z \gamma(\to \mu^+ \mu^-)}

\newcommand\ks{K^0_S }

\newcommand\dm{X17}
\newcommand\mee{m_{e^+e^-}}
\newcommand\angee{\Theta_{ e^+e^-}}

\begin{document}

\title{Hunting down the X17 boson at the CERN SPS} 

\subtitle{NA64 Collaboration}
\author{
E.~Depero\thanksref{addrETH} \and                     
Yu.~M.~Andreev\thanksref{addrINR} \and                
D.~Banerjee\thanksref{addrCERN,addrUIL} \and          
J.~Bernhard\thanksref{addrCERN} \and                  
V.~E.~Burtsev\thanksref{addrJINR} \and                
N.~Charitonidis\thanksref{addrCERN} \and              
A.~G.~Chumakov\thanksref{addrTGPU} \and               
D.~Cooke\thanksref{addrUCL} \and                      
P.~Crivelli\thanksref{addrETH,pc}                     
A.~V.~Dermenev\thanksref{addrINR} \and                
S.~V.~Donskov\thanksref{addrIHEP} \and                
R.~R.~Dusaev\thanksref{addrTPU} \and                  
T.~Enik\thanksref{addrJINR} \and                      
A.~Feshchenko\thanksref{addrJINR} \and                
V.~N.~Frolov\thanksref{addrJINR} \and                 
A.~Gardikiotis\thanksref{addrUPT} \and                
S.~G.~Gerassimov\thanksref{addrTUM,addrLPI} \and      
S.~Girod\thanksref{addrCERN} \and                     
S.~N.~Gninenko\thanksref{addrINR} \and                
M.~H\"osgen\thanksref{addrUBN} \and                   
V.~A.~Kachanov\thanksref{addrIHEP} \and               
A.~E.~Karneyeu\thanksref{addrINR} \and                
G.~Kekelidze\thanksref{addrJINR} \and                 
B.~Ketzer\thanksref{addrUBN} \and                     
D.~V.~Kirpichnikov\thanksref{addrINR} \and            
M.~M.~Kirsanov\thanksref{addrINR} \and                
V.~N.~Kolosov\thanksref{addrIHEP} \and                
I.~V.~Konorov\thanksref{addrTUM,addrLPI} \and         
S.~G.~Kovalenko\thanksref{addrUNAB} \and              
V.~A.~Kramarenko\thanksref{addrJINR,addrMSU} \and     
L.~V.~Kravchuk\thanksref{addrINR} \and                
N.~V.~Krasnikov\thanksref{addrJINR,addrINR} \and       
S.~V.~Kuleshov\thanksref{addrUNAB} \and               
V.~E.~Lyubovitskij\thanksref{addrTGPU,addrUSM} \and   
V.~Lysan\thanksref{addrJINR} \and                     
V.~A.~Matveev\thanksref{addrJINR} \and                
Yu.~V.~Mikhailov\thanksref{addrIHEP} \and             
L.~Molina~Bueno\thanksref{addrETH} \and               
D.~V.~Peshekhonov\thanksref{addrJINR} \and            
V.~A.~Polyakov\thanksref{addrIHEP} \and               
B.~Radics\thanksref{addrETH} \and                     
R.~Rojas\thanksref{addrUSM} \and                      
A.~Rubbia\thanksref{addrETH} \and                     
V.~D.~Samoylenko\thanksref{addrIHEP} \and             
D.~Shchukin\thanksref{addrLPI} \and                   
H.~Sieber\thanksref{addrETH} \and                     
V.~O.~Tikhomirov\thanksref{addrLPI} \and              
vI.~Tlisova\thanksref{addrINR} \and                    
D.~A.~Tlisov\thanksref{dt,addrINR} \and               
A.~N.~Toropin\thanksref{addrINR} \and                 
A.~Yu.~Trifonov\thanksref{addrTGPU} \and              
B.~I.~Vasilishin\thanksref{addrTPU} \and              
G.~Vasquez Arenas\thanksref{addrUSM} \and             
P.~V.~Volkov\thanksref{addrJINR,addrMSU} \and         
V.~Yu.~Volkov\thanksref{addrMSU} \and                 
P.~Ulloa\thanksref{addrUNAB}                          
}

\thankstext{pc}{Corresponding author, e-mail: Paolo.Crivelli@cern.ch} 
\thankstext{dt}{Deceased}

\institute{
ETH Z\"urich Institute for Particle Physics and Astrophysics,
CH-8093 Z\"urich, Switzerland\label{addrETH}                                    
\and 
Institute for Nuclear Research, 117312 Moscow, Russia\label{addrINR}            
\and 
CERN, EN-EA, 1211 Geneva 23, Switzerland\label{addrCERN}                        
\and
University of Illinois at Urbana Champaign, Urbana,
61801-3080 Illinois, USA\label{addrUIL}                                         
\and 
Joint Institute for Nuclear Research, 141980 Dubna, Russia\label{addrJINR}      
\and 
Tomsk State Pedagogical University, 634061 Tomsk, Russia\label{addrTGPU}        
\and 
UCL Departement of Physics and Astronomy, University College London,             
Gower St. London WC1E 6BT, United Kingdom\label{addrUCL}                        
\and 
State Scientific Center of the Russian Federation Institute for High Energy 
Physics of National Research Center 'Kurchatov Institute' (IHEP), 
142281 Protvino, Russia\label{addrIHEP}                                         
\and 
Tomsk Polytechnic University, 634050 Tomsk, Russia\label{addrTPU}               
\and 
Physics Department, University of Patras, 265 04 Patras, 
Greece\label{addrUPT}                                                           
\and 
Technische Universit\"at M\"unchen, Physik  Department, 85748 Garching, 
Germany\label{addrTUM}                                                          
\and  
P.N. Lebedev Physical Institute, 119 991 Moscow, Russia\label{addrLPI}          
\and  
Universit\"at Bonn, Helmholtz-Institut f\"ur Strahlen-und Kernphysik, 
53115 Bonn, Germany\label{addrUBN}                                              
\and 
Departamento de Ciencias F\'{i}sicas, Universidad Andres Bello, Sazi\'{e} 2212, 
Piso 7, Santiago, Chile\label{addrUNAB}                                         
\and 
Skobeltsyn Institute of Nuclear Physics, Lomonosov Moscow State University, 
119991  Moscow, Russia\label{addrMSU}                                           
\and 
Universidad T\'{e}cnica Federico Santa Mar\'{i}a, 2390123 Valpara\'{i}so, 
Chile \label{addrUSM}                                                           
}

\maketitle

\begin{abstract}
Recently, the ATOMKI experiment has reported new evidence for the excess of $\pair$ events 
with a mass $\sim$17 MeV in the nuclear transitions of $^4$He, that they previously observed 
in measurements with $^8$Be. These observations could be explained by the existence of 
a new vector $\dm$ boson. So far, the search for the decay $\xdecay$  with the NA64 experiment 
at the CERN SPS gave negative results. Here, we present a new technique that could be implemented 
in NA64 aiming to improve the sensitivity and to cover the remaining $\dm$ parameter space. 
If a signal-like event is detected, an unambiguous observation is achieved by reconstructing 
the invariant mass of the $\dm$ decay with the proposed method. To reach this goal 
an optimization of the $\dm$ production target, as well as an efficient and accurate 
reconstruction of two close decay tracks, is required. A dedicated analysis of the 
available experimental data making use of the trackers information is presented. 
This method provides independent confirmation of the NA64 published results \cite{visible-2018-analysis},
validating the tracking procedure.
The detailed Monte Carlo study of the proposed setup and 
the background estimate show that the goal of the proposed search is feasible.
\end{abstract}

\section{Introduction}
\label{sec:introduction}
Dark sectors are very interesting candidates to explain the origin of Dark Matter 
(see, e.g., Ref.~\cite{mb} for a recent review), whose presence has 
so far been inferred only through its gravitational interaction from cosmological observations \cite{hooper}. 
If, in addition to gravity, a new force between the dark sector and visible matter exists \cite{prw, pospelov} this can be tested in laboratory experiments. A possibility is that this new force is carried by a vector boson $A'$, called dark photon.
Stringent limits on the coupling strength $\epsilon$ and mass $m_{A'}$ of such dark photons, excluding the parameter space region favored by the anomalous magnetic moment of the muon, the so called $(g-2)_{\mu}$ anomaly, have already been placed by beam dump \cite{jdb, charm, rio, e137, konaka, bross, dav,  ath, nomad, Adler:2004hp, essig1, blum,sg1, blum1, sarah1}, fixed target \cite{apex,merkel,merkel1}, collider \cite{babar, curt, babar1}, rare particle decay searches \cite{sindrum, kloe, sg2, kloe2, wasa, hades, phenix, e949, na48, Dubinina:2017smx, kloe3} and the new determination of the fine structure constant $\alpha$ combined with the measurement of $(g-2)_e$ \cite{Parker191,PhysRevLett.100.120801}. 

A great boost to search for the new light boson weakly coupled to Standard Model particles was triggered by the recent observation of a $\sim$7$\sigma$ excess of events in the angular distribution of $\pair$ pairs produced in the nuclear transitions of the excited $^8$Be$^*$ nuclei to its ground state via internal $\ee$ pair creation \cite{be8,be8-1}. The latest results of the ATOMKI group report a similar excess at approximately the same invariant mass in the nuclear transitions of another nucleus, $^4$He \cite{be8-2}.

It was put forward  \cite{feng1,feng2}, that this anomaly can be interpreted as the emission of a protophobic gauge boson $\dm$ decaying into $\pair$ pairs. To be consistent with the existing constraints, the $\dm$ boson should have a non-universal coupling to quarks and a coupling strength with electrons in the range of $2\times 10^{-4} \lesssim \epsilon \lesssim 1.4\times 10^{-3}$ which translates to a lifetime of the order of $10^{-14}\lesssim \tau_X \lesssim 10^{-12}$~s. Remarkably, this model also explains within experimental uncertainty the new result obtained with the $^4$He nucleus, providing both kinematical and dynamical evidence to support this interpretation \cite{feng3}. Recently, this explanation was challenged as it would also imply $\dm$ production via bremsstrahlung radiation \cite{zhang2020protophobic}, not observed in the experiment. The NA64 collaboration aims to probe the $\dm$ in a model independent way. Our setup is sensitive to the full set of couplings estimated to be in the range $10^{-5} \lesssim \epsilon \lesssim 1.4\times 10^{-3}$, which could explain the anomaly for the scalar, pseudo-scalar, vector and axial-vector cases \cite{x17-plans-na64}. However, in this work, we present the NA64 current results and the projected sensitivity of the future new setup using a protophobic vector boson as benchmark model.

Interestingly, such a new boson with a relatively large coupling to charged leptons could also resolve the tension between measured and predicted values of the $(g - 2)_{\mu}$. In addition to vector and axial-vector explanation of the $\dm$ anomaly, one can consider scenarios involving hidden pseudo-scalar boson \cite{Ellwanger:2016wfe}. Corresponding pseudo-scalar couplings to electrons satisfy existing experimental constraints \cite{Andreas:2010ms,Adler:2004hp}. An analysis to probe such pseudo-scalar states at NA64 \cite{Kirpichnikov:2020tcf} would require a proper Monte-Carlo simulation of the spectra and flux of light pseudo-scalar boson produced in the target by electrons. This code is currently under developement, and it will allow us to use data collected in our previous analysis \cite{Banerjee:2020fue} to probe values of coupling in the region $10^{-5} \lesssim \epsilon \lesssim 10^{-4}$, also cross-checking the region of parameter space already covered by E141 \cite{blum}.
Another interesting result comes from the new measurement of $\alpha$ performed by Parker et al. \cite{Parker191} which combined with the $(g-2)_e$ measurements results in a 2.4$\sigma$ deviation from the QED predictions \cite{PhysRevLett.100.120801}. Should this tension be confirmed by the planned improvement of Parker's et al. measurements, the two constraints coming from the NA64 results and $(g - 2)_e$ would exclude the vector and axial vector couplings explanation of $\dm$. On the other hand, models with nonzero V$\pm$A coupling constant with the electron would explain both electron and muon $(g - 2)$ anomalies \cite{Krasnikov:2019dgh}. In these models, the $\dm$ could have a coupling of $6.8\cdot 10^{-4} \lesssim \epsilon \lesssim 9.6 \cdot 10^{-4}$ which leaves an interesting region of the parameter space to be explored.
These models motivated the study of the phenomenological aspects of such a light vector boson weakly coupled to quarks and leptons (see, e.g., Refs.~\cite{fayet1, fayet2, fayet3, fayet4,jk, cheng, Zhang:2017zap, ia, liang, bart}) 
and new experimental searches (see, e.g., Refs.~\cite{mb, nardi}).

Recently, the NA64 collaboration has reported new results that excluded the $\dm$ boson  with the coupling strength  to electrons in the range $1.2 \times 10^{-4} < \epsilon < 6.8 \times 10^{-4}$ \cite{NA64Be2017,visible-2018-analysis}, by using the calorimeter technique proposed in \cite{Gninenko:2013rka,Andreas:2013lya}. In this work, the main challenges to search for large coupling $\epsilon \sim 10^{-3}$ of $\dm$ will be outlined and an upgrade of the setup to overcome them is described. First, in Sec.\ref{sec:2018-setup} an overview on the calorimeter method \cite{Gninenko:2013rka,Andreas:2013lya,visible-2018-analysis} is presented and the main limitations of the current setup are outlined. In Sec.\ref{sec:new-tracking-analysis}, a new analysis method that exploits the trackers is presented. This analysis highlights the importance of an efficient tracking procedure for the $\dm$ search. The increase in sensitivity is however negligible due to the intrinsic limitations of the setup. In Sec.\ref{sec:new-visible-setup} a new setup optimized for searching the $\dm$ and $A'$ with large couplings $\epsilon \sim 10^{-3}$ is described. The method for the invariant mass reconstruction of $\ee$ pairs is presented in Sec.\ref{sec:imassreco}. Our conclusions are reported in Sec.\ref{sec:conclusion}.

\section{2018 Visible mode setup}
\label{sec:2018-setup}

The method of the search for $\aee$ (or $\xdecay$) decays is detailed in \cite{Gninenko:2013rka, Andreas:2013lya, gkkk1, DMsimulation}. Here, we review it briefly. The $\dm$ is produced via scattering of 150 GeV electrons off nuclei of an active target-dump. The $\dm$ production is followed by its decay into $\ee$ pairs:
\begin{equation}
e^- + Z \to e^- + Z + \dm (\to \ee) \,.
\label{ea}
\end{equation}

\begin{figure*}[tb]
\centering
\includegraphics[width=1.\textwidth]{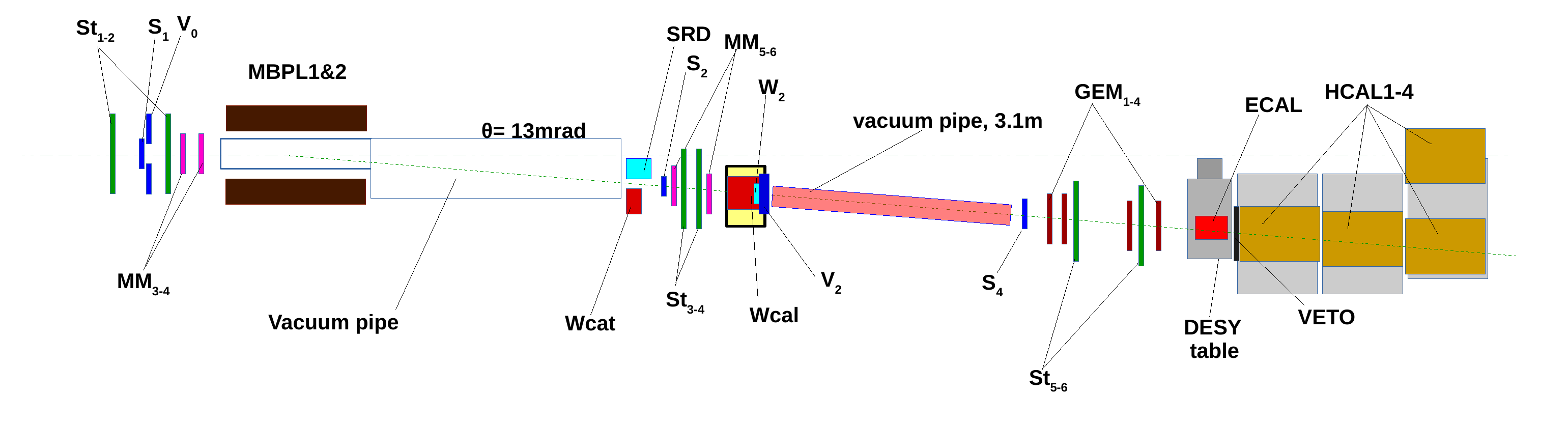}
\caption{The setup used in 2018 to search for $\dm$$\to \ee$  decays of the bremsstrahlung $\dm$ produced in the reaction
$eZ \to eZ+\dm $ of the 150 GeV electrons incident on the active WCAL target.}
\label{fig:setup-2018}
\end{figure*}

The NA64 experiment searched for these decays using the H4 beam line of the CERN Super Proton Synchrotron (SPS) / North Area, delivering $\simeq 5\times 10^6~e^-$ every $\sim$30 seconds with an average spill length of 4.8 seconds. The setup used for this search is shown in Fig.\ref{fig:setup-2018}. A spectrometer made of two bending magnets (MBPL) combined with Micromegas tracker chambers (MM) measures the particle momentum \cite{na64-micromegas}. The magnet allows as well a very efficient electron identification (ID) using a Synchrotron Radiation Detector(SRD) segmented in three different counters \cite{Depero:2017mrr}. Two large area (20 cm$^2$) Strawtubes (St) \cite{Volkov:2019qhb,Dermenev:2015ehab} are placed after the vacuum tube to detect particles with large divergence from the beam originating for example from charged hadronic secondaries produced in the vacuum window and then scattering at large angles. The active target used for the conversion is sandwich electromagnetic-calorimeter (WCAL) made of tungsten and scintillator layers, the longitudinal dimension is minimized to boost the probability of the $\dm$ of decaying outside the dump. The WCAL is longitudinally segmented in a pre-shower part ($\simeq$5$X_0$) used to suppress the background coming from hadrons and a calorimeter part ($\simeq$25$X_0$) to completely stop incoming electrons. Hadrons are additionally suppressed using a high efficiency VETO and a set of 3 hadronic calorimeter modules (HCAL) placed at the end of the setup. To measure the energy of the $\ee$ in signal events a hodoscopic electromagnetic-calorimeter (ECAL) is placed downstream of the decay volume. The signal region is defined by the sum of energy deposited in both calorimeters being compatible to the original beam energy. Additionally, the energy deposited in the last layer of the WCAL (W$_2$) is required to be lower than one deposited by a Minimum Ionizing Particle (MIP) to suppress punch-through secondaries from the electromagnetic or hadronic shower in the target. The presence of an $\ee$ pair in the decay volume is assessed by a scintillator counter placed immediately after the decay volume (S4). The trigger used required in-time energy deposition in the S$_{1-3}$ counter, no energy deposition in V$_0$ and $E_{WCAL} \lesssim 0.7 \times E_{beam}$ \cite{visible-2018-analysis}.

The allowed coupling $\epsilon$ for the $\dm$ can be as high as $1.4 \times 10^{-3}$, resulting in its very short decay length of few mm. Therefore, to boost the signal yield one should reduce the length of the active target to enhance the number of decays outside the dump. Additionally, for an unambiguous signature of the $\dm$ production it is crucial to reconstruct the invariant mass of the $\ee$ pair. Their small opening angle of $\angee \lesssim$0.3 mrad makes this task particularly challenging. The method proposed to solve this problem is discussed in Sec.\ref{sec:imassreco}.

\section{2018 visible mode analysis using trackers}
\label{sec:new-tracking-analysis}

The published analysis using the data collected in 2018 \cite{visible-2018-analysis} was based exclusively on the calorimetry approach discussed in Sec.\ref{sec:2018-setup}. The 4 Gaseous Electron Multiplier (GEM) trackers after the decay volume were not used for the signal discrimination. Here we present a novel method which exploits them providing a boost in the signal yields while maintaining the background under control. Even though this analysis is not sensitive when the decay length of the $\dm$ is significantly smaller than the dimension of the dump, it has the advantage of being complementary to the calorimeter analysis. Moreover, it is a very important proof of principle to demonstrate the power of our tracking procedure.

\begin{figure*}[tbh!]
  \centering
  \includegraphics[scale=0.9]{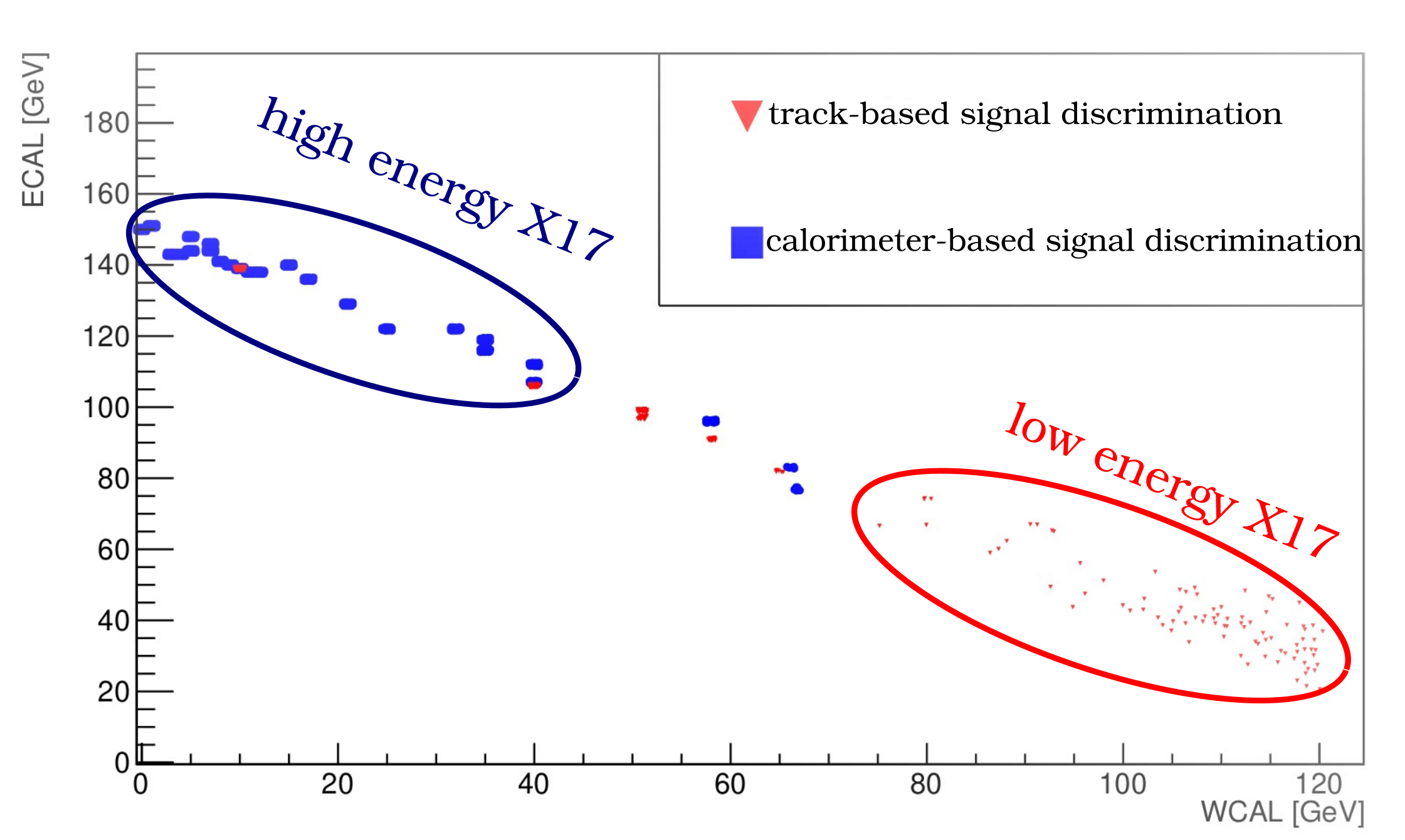}
  \caption[combined-analysis]{$\dm$ simulated in the visible mode 2018 setup. Two different cuts are used to discriminate between two $\dm$ topologies. The first one is based on angle cut and vertex position using information from the 4 GEM stations installed in the decay volume and is very efficient on the $\dm$ produced at low energy (red triangle). The second one relies on the Veto placed at the end of the dump and is more efficient for the high energy population (blue square).}
  \label{fig:combined-analysis}
\end{figure*}

While in a first approximation the $\dm$ is produced in the first few layers of the WCAL, it can also originate at a later stage of the em-shower. These events, which are typically rejected in the calorimeter analysis, are instead accepted in the new analysis presented here. First, an initial sample is selected in the same way described in Sec.\ref{sec:2018-setup} and detailed in \cite{visible-2018-analysis}. The final discrimination in the calorimeter analysis is based on the counter W2 placed at the end of the dump to reject the charged punch-through from the em-shower. This last cut is efficient if the $\dm$ is produced in the first few layers of the WCAL, but typically reject the event if the $\dm$ is produced at a later stage of the em-shower. The reason is that these events are accompanied by a long longitudinal development of the em-shower that leaves an energy deposit larger than the typical energy cut accepted in the calorimeter analysis. On the other hand, the low energy of the produced $\dm$ implies a larger angle between the decay products that can be resolved by the trackers. Combining these two concepts, one can see that the signal yield is characterized by two different topologies that can be easily distinguished by looking at the energy deposited in the ECAL (see Fig.\ref{fig:combined-analysis}).

Using this distinction, we divide all events that passed the initial selection criteria in two topologies based on the total energy deposited in the ECAL. The exact value of this threshold was selected to maximize the signal yield. The optimal value has a small dependence on the $\dm$ mass and coupling. A simple threshold of 75 GeV amounting to half of the initial beam energy was found to be robust for most of the interesting signal scenario. After the topology is decided, a final set of cuts is applied to discriminate between signal and background. In the case of high energy $\dm$, trackers do not have the capability of discriminate between single hits. An energy deposit smaller than 0.8 E$_{MIP}$ is required in W2, and the presence of a decay after the dump is assessed by asking S4 (see Fig.\ref{fig:setup-2018}) to have an energy deposited larger than 1.5 E$_{MIP}$. On the other hand, if the energy deposited in the ECAL is smaller than 75 GeV, trackers are used instead as final discriminator. Two tracks in the decay volume are required with a reconstructed vertex within 3$\sigma$ from the WCAL and an angle smaller than 3 mrad. This different treatment leads to an increased efficiency to $\dm$ produced at a late stage of the shower as shown in Fig.\ref{fig:combined-analysis}. However, the smaller energy of the $\dm$ produced in this way has the effect of reducing the probability of the particle escaping the dump. For large coupling $\epsilon$ this suppression can be more than 2 orders of magnitude, making the boost of signal yield negligible. A summary of this boost for various interesting $\dm$ and $A'$ scenario is illustrated in Table \ref{tab:dm:efftable}. The values reported consider also a conservative correction factor of 0.77$\pm$0.1 that takes into account inefficiencies of the detectors and the reconstruction algorithm. This factor was evaluated using a data-driven method precisely outlined in Sec.\ref{sec:study-dimu-prod}. The conclusion of this study is that in the current setup trackers information do not improve the limit on the $\dm$ parameter space. This is because the boost in signal yield becomes negligible for $\epsilon \sim 6 \times 10^{-4}$, a value which is already excluded with 90\% confidence by our previous analysis.

\begin{center}
\begin{table}
  \centering
  \begin{tabular}{|l|l|r|}
    \hline
    M$_{A'}$ [GeV]& $\epsilon$ & N$^{new}_{A'}$ / N$^{old}_{A'}$ \\    
    \hline
    0.005  & 0.004    & 1   \\    
    0.01   & 0.0015   & 1   \\    
    0.01   & 0.003    & 1   \\    
    0.0167 & 0.0001   & 1.22\\
    0.0167 & 0.00018  & 1.2 \\    
    0.0167 & 0.000316 & 1.2 \\
    0.0167 & 0.0006   & 1.01\\
    0.0167 & 0.0007   & 1   \\
    0.022  & 0.000316 & 1.22\\
    \hline    
  \end{tabular}
  \caption{N$^{new}_{A'}$ / N$^{old}_{A'}$ ratio between signal events observed in tracker-analysis compared to calorimeter-only analysis. The new analysis uses cuts based on GEM tracking detectors if the energy detected by the downstream ECAL is below 75 GeV.}
  \label{tab:dm:efftable}
\end{table}
\end{center}

\subsection{Study of dimuon production in 2018 setup}
\label{sec:study-dimu-prod}

To validate the MC simulation and the tracking procedure required for this analysis, a pure sample of events produced in the WCAL from the rare QED interaction $\emu$ has been studied. This class of events has many similarities to the $\dm$ ones, and they can be easily selected by requiring a double MIP signature in the HCAL modules. This procedure is described in detail in \cite{na64-prd}. The double tracks expected in the decay volume are then used to test the reliability of the tracking procedure in the setup.

\begin{figure*}[tbh!]
  \begin{center}
    \includegraphics[scale=0.8]{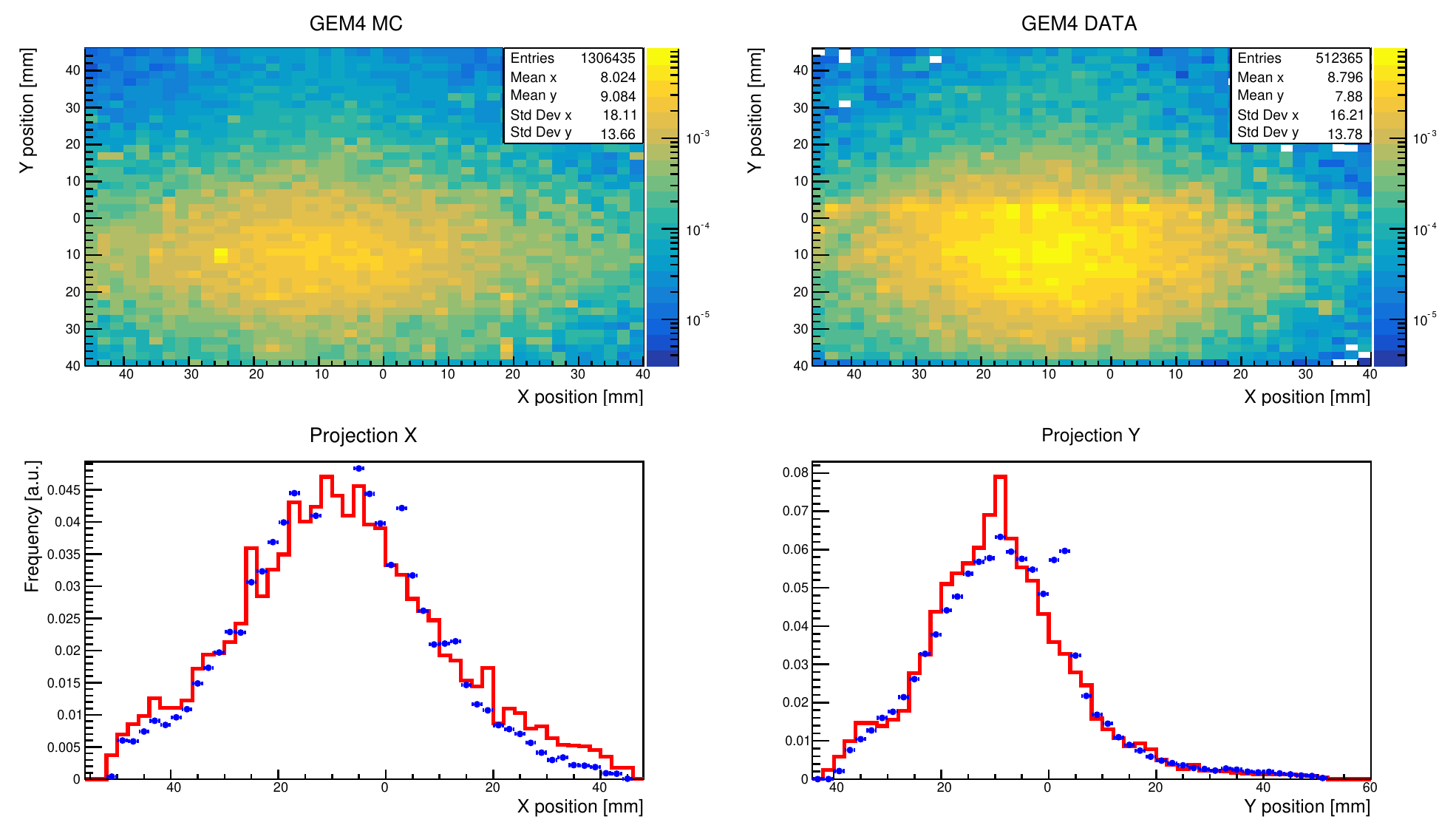}
  \end{center}
  \caption{Hit position recorded in last GEM before ECAL for MC simulated (Red curve) and data (Blue dots) $\emu$ events.}
  \label{fig:dimuon:gemspectra}
\end{figure*}

To improve the quality of the MC, a realistic beam profile was extracted from the electron calibration runs. Hadrons in the sample were rejected by requiring an energy between 5 MeV and 100 MeV for both SRD counters. The beam profile is then calculated by fitting the XY position recorded by MM3,4 in Fig.\ref{fig:setup-2018} with a 2D Gaussian. The two fits agree within 100 $\mu$m precision for both $\sigma_x \approx 4.13$ mm and $\sigma_y \approx 1.40$ mm. Fig.\ref{fig:dimuon:gemspectra} shows a comparison of the reconstructed hit-position between data and MC in $\emu$ events after the extracted beam profile is used in the simulation.

To further improve the agreement between data and MC several strategies were used. The limited spatial resolution of GEMs was taken into account by applying a smearing of 80 $\mu$m. This number was estimated by checking track residuals after selecting different GEMs triplets for the track reconstruction and comparing the reconstructed hit to the one predicted by the tracking procedure. To reproduce the single planes of the GEM, hits are separated in X-Y projections and knowledge on the original hit combination is no longer assumed. As the minimal hit separation between hits in GEM was conservatively estimated to be 1.75 mm, hits closer than this threshold were merged in the MC. This number was estimated using clusters recorded by the GEM detectors during electron run in 2017. The hits generated in this procedure are used as input for the same reconstruction algorithm used for the data.

The reconstruction chain works as follows:
\begin{enumerate}
\item Track candidates are defined by grouping hits where the angle between first and second GEM pair is smaller than 9 mrad.
\item Those candidates are reconstructed using a Kalman filter implemented with the Genfit library \cite{genfit}.
\item Vertex candidates are generated by grouping tracks pair with no common hits.
\item The exact position of the vertex is obtained by back-propagating the tracks at their point of minimum distance. Only vertices with a distance below 3 mm are considered for the analysis.
\end{enumerate}

A dimuon sample was selected using all events collected during the visible mode 2018 run (3$\times 10^{10}$ EOT). The beam quality was improved by requiring the reconstructed momentum to be in the range between 140 and 160 GeV. The $\emu$ events leave a double-MIP signature in each HCAL module, thus a cut 2 GeV$<$ E$_{hcal} <$ 6.35 GeV is applied for the selection. Since an hardware trigger which selects only events with missing energy in the WCAL is used during the data taking, an additional cut $E_{WCAL} < 90$ GeV is applied to consider only such events in both simulation and data. This cut also selects a sample with kinematics closer to the one expected from a $\dm$ candidate. This makes the comparison with the MC more significant for our search. Scintillator counters also need to be compatible with a $\mu^+ \mu^-$ in the decay volume: an energy deposited of at least 1 E$_{MIP}$ is required in the scintillator (S4) downstream the WCAL and at least 1.8 E$_{MIP}$ in the Veto behind the ECAL. The less stringent cut on S4 is justified by its limited transverse dimension which makes it not suitable for a precise energy measurement.

Although these cuts mainly select dimuon generated from $e^-$ primaries, a contribution is also expected from the hadron contamination. The physical trigger employed in the experiment further increases such contribution, as the requirement of low energy deposit in the WCAL bias the beam composition to particles with high penetration power. To solve this issue, a cut on the SRD detector and on the WCAL pre-shower are used. These cuts are expected to reject hadrons and muons at a level $<10^{-5}$.

\begin{figure*}[tbh!]
  \begin{center}
    \includegraphics[scale=0.8]{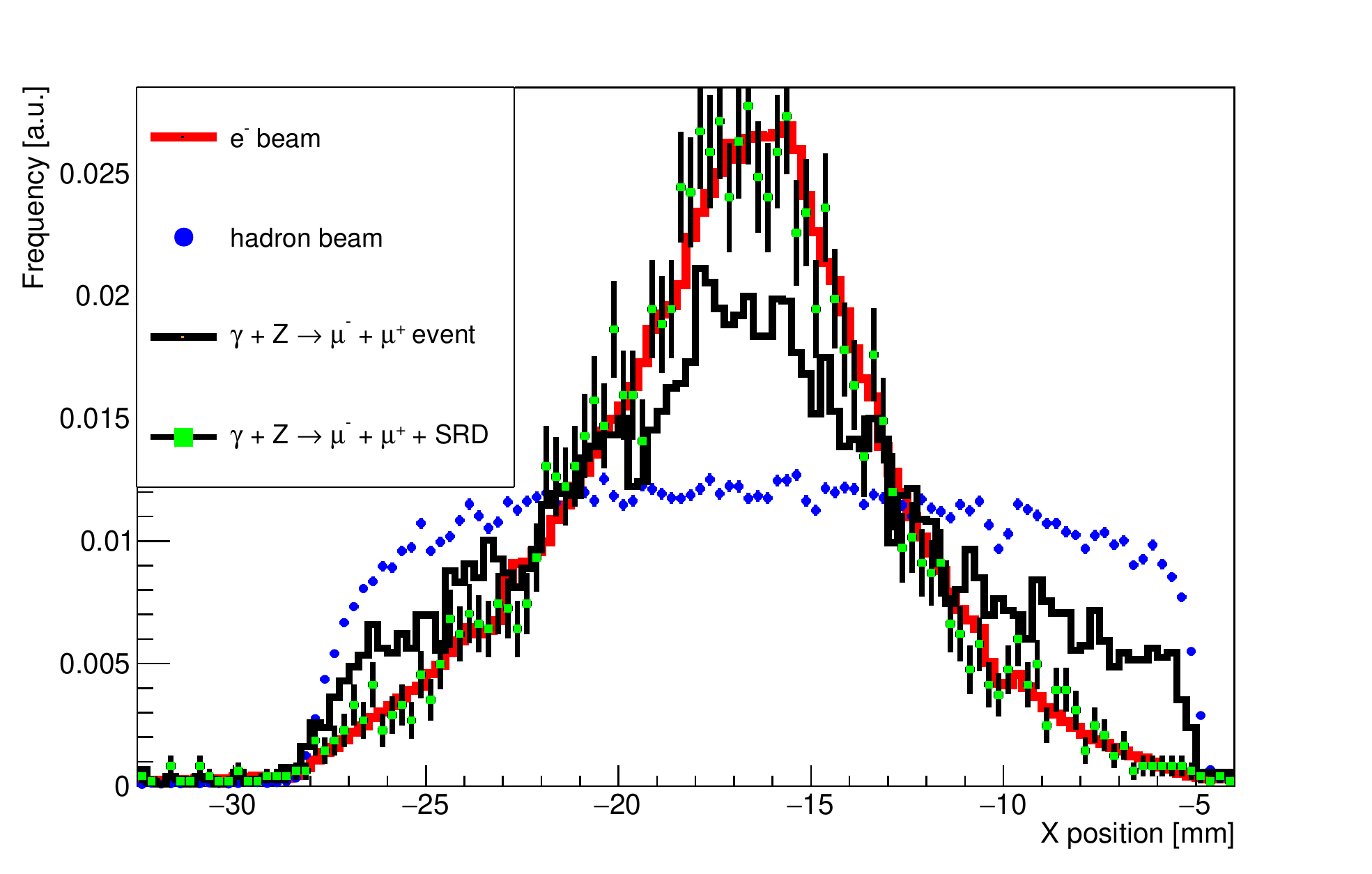}
  \end{center}
  \caption{Beam profile recorded during 2018 data taking by the first Micromegas module upstream for hadron calibration run (blue dots), electron calibration run (red line), events selected with dimuons cuts from data collected with the physical trigger (black line) and those same events after SRD cut is applied (green square). Fits using the templates obtained from the calibration run show a level of contamination of $\sim$50\% in the dimuon sample. The contamination is completely removed after the SRD cut is applied.}
  \label{fig:dimuon:profile}
\end{figure*}

To cross-check that the contamination is correctly removed, an independent method based on the beam profile shape is used. The beam profile significantly differs between electrons and hadrons as the H4 beamline is tuned for selecting electrons in our search. Both profiles are recovered from the data using a calibration run of electron/hadron respectively. Using a  $\chi^2$-test the ratio between the two is estimated by mixing the two templates until the best agreement with the measured beam profile is reached. The result is summarized in Fig.\ref{fig:dimuon:profile}: the beam profile of dimuon-selection events is compared before and after the SRD criteria is applied. The fit shows a contamination of roughly 50\% in the original sample. After the cut the beam profile converges to the templates obtained in the $e^-$ calibration runs, resulting in an estimated contamination level $<$1\%.

\begin{figure*}[tbh!]
  \begin{center}
    \includegraphics[scale=0.8]{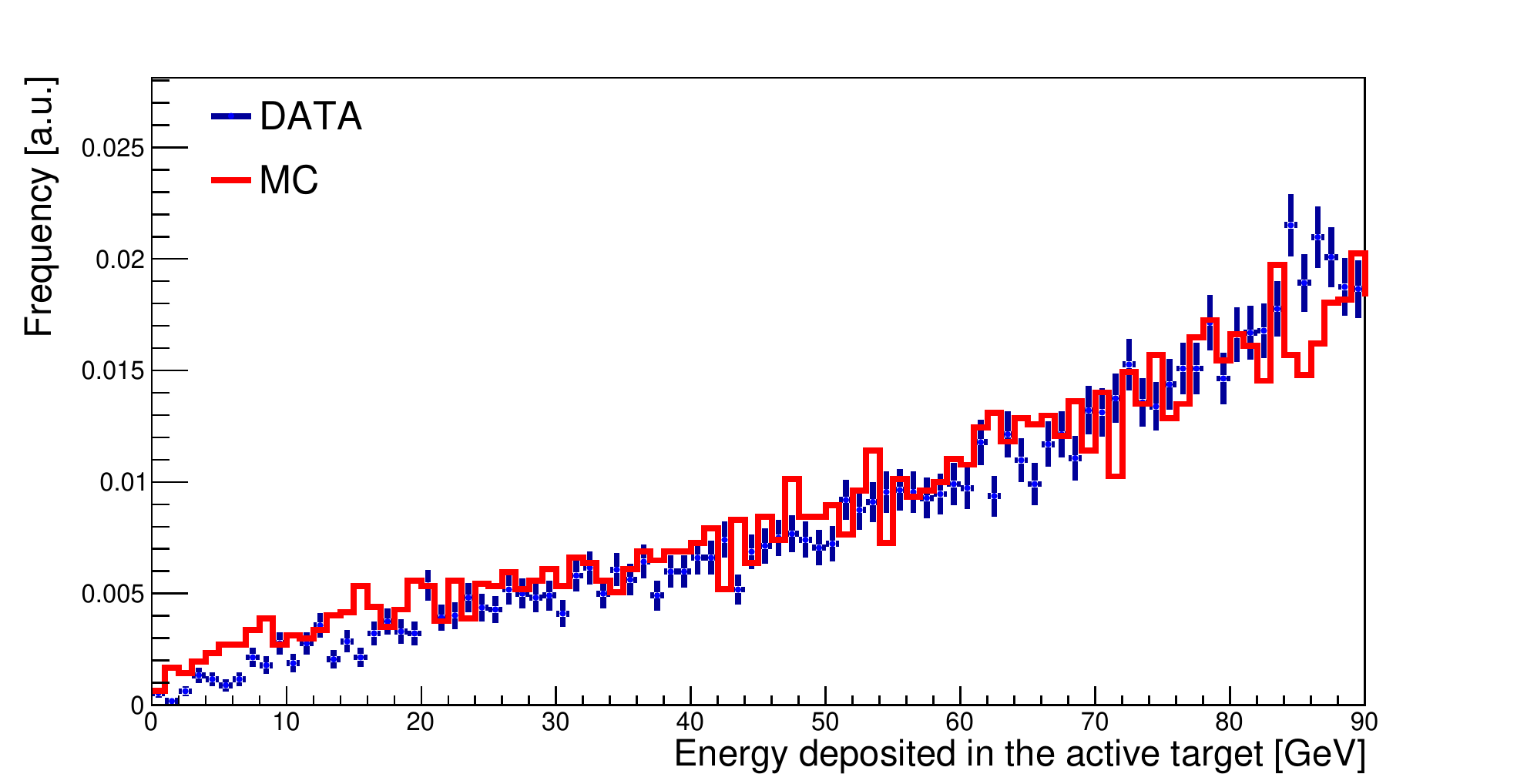}    
  \end{center}
  \caption{Energy deposit in the active dump (WCAL) after all selection criteria are applied in $\emu$ event for data collected in 2018 (blue dots) and MC-generated events (red line). }
  \label{fig:dimuon_en}
\end{figure*}

To show that there are no significant differences in the tracking procedure between simulation and data the energy deposited in the WCAL was used as a figure of merit. If the tracking procedure affects differently data and MC, one would expect the agreement between the two distributions to diverge after applying cuts based on vertex reconstruction. Following the procedure described above, a number of vertex candidates are selected for the comparison. As the interaction $\emu$ will have their vertex inside the WCAL, only vertices compatible with this assumption are selected for the comparison. In practice, a vertex is accepted if its position lies within 3$\sigma$ of the expected WCAL position, where $\sigma$ was fitted using a Gaussian from the distribution of $\mu^- \mu^+$ pairs selected from the simulation. After the selection criteria, the distributions of energy deposited in the WCAL obtained for MC and data samples are compared as shown in Fig.\ref{fig:dimuon_en}. One can see that the distributions are in excellent agreement, thus proving that the tracking cuts do not bias the original sample.

A lower efficiency is observed in the data compared to the Monte Carlo after event selection. The reasons for this are inefficiency of the GEM modules, fail of clusterization in some events and differences in the tracking procedure due to the simplifications used in the MC. The cuts applied to the sample are divided into four steps. First, at least two hits per GEM are required in the decay volume as a minimal condition for tracking. After that, events with a GEM module recording more than 5 hits are rejected as incompatible with a single $\emu$ vertex. The MC predicts 68\% of $\mu^+ \mu^-$ pair after these two cuts. This low acceptance is caused by the GEMs position optimized to resolve very close hits coming from the decay of $\dm$. These selection criteria do not depend on the track-fitting procedure but instead rely on the clusterization performed and the efficiency of the trackers. Tracking procedure is then applied to the events that survived the two first requirements. The reconstructed vertex position is required to be compatible with a vertex inside the dump. The number of events surviving the last requirement  is slightly smaller in the data. The disagreement between the ratio of good vertices reconstructed inside the decay volume is $<$1\%. Finally, a factor of 0.77 estimated from the analysis of data and MC samples accounting for all these differences, is used to correct the 2018 signal yield. A summary of the efficiency can be found in Table \ref{tab:dimuon:efficiencies}.

\begin{center}
\begin{table*}
  \centering
  \begin{tabular}{|l|c|c|c|}
    \hline
    Cut & Efficiency MC & Efficiency Data & MC / DATA \\
    \hline
    \textbf{Hit} & & &\\
    \hline
    Hits per GEM $\geq$ 2 & 0.68$\pm$0.1 & 0.58$\pm$0.1 & 0.85$\pm$0.1 \\
    Hits per GEM $\leq$ 5 & 0.68$\pm$0.1 & 0.55$\pm$0.1 & 0.80$\pm$0.1 \\
    \hline
    \textbf{tracking} & & &\\
    \hline
    Vertex distance $\leq$ 3 mm & 0.63$\pm$0.1 & 0.49$\pm$0.1 & 0.77$\pm$0.1  \\
    Vertex in decay volume & 0.62$\pm$0.1 & 0.48$\pm$0.1 & \textbf{0.77$\pm$0.1}\\
    \hline
    
  \end{tabular}
  \caption{Efficiency of cuts based on tracking criteria for a clean sample of simulated $\emu$ and dimuon selected from 2018 data. The efficiency presented in the table are cumulative, with the first cut applied being the one in the first row. First two cuts are based exclusively on information coming from the single GEM modules. Last two cuts are based on the tracking procedure.}
  \label{tab:dimuon:efficiencies}
\end{table*}
\end{center}

\label{sec:methods:comparison}

\subsection{Background estimate in visible mode analysis}
\label{sec:new-analysis:background}

Background for tracking based approach could arise from particles punching through the WCAL and leaving a signature in the trackers downstream. The main sources of this are either large energy $\gamma$ not interacting in the WCAL and converting in the last few layers or hadrons interacting in the dump.

In the case of hadrons, inelastic scattering in the WCAL produces a large occupancy in the decay volume that can potentially create vertex candidates. Such events are expected to be suppressed by the selection criteria applied downstream for hadron rejection outlined in Sec.\ref{sec:new-tracking-analysis}. Furthermore, events able to mimic the pure electromagnetic signal of the decay $\dm \rightarrow e^+ e^-$ are often accompanied by a large transversal spread and are thus rejected by the requirement of energy conservation at a level of $< 10^{-5}$. This estimate was obtained by integrating the events in the signal region with two tracks in the decay volume without applying any rejection criteria for hadrons in a $\pi^-$ simulation. Such an event should also pass the independent selection criteria applied upstream, namely large energy deposited in the SRD and WCAL pre-shower. In a sample up to $10^7$ EOT, it was not possible to find an event with such signature even after removing the HCAL and VETO from the selection criteria. It can be concluded that this background is negligible for the EOT accumulated during 2018.

To estimate the background for a larger number of EOT, the $\ks$ decay was used as a benchmark process, as its short decay length is expected to be compatible with the ones of the $\dm$. The energy spectrum of $\ks$ was simulated using an exponential distribution with an energy cut-off of 18 GeV, as $\ks$ below this energy have a negligible probability to decay outside the dump. By applying tracking-criteria over this sample it was estimated that a rejection of $10^{-2}$ can be conservatively achieved for this background using the opening angle of the reconstructed vertex as discriminator. This estimate however mostly depends on the main hadronic decay channel $\ks \rightarrow \pi^- + \pi^+$ which is further suppressed downstream by the hadron suppression cuts such as no energy deposited in the HCAL and in the VETO (see Fig.\ref{fig:setup-2018}). The decay channel $K^0_S \rightarrow \pi^0 + \pi^0$ has on the other hand a small chance to leave any signature in the GEM modules as no charged particle is typically emitted. Signal-like events can be produced either by the conversion of a photon from the $\pi^0 \rightarrow \gamma \gamma$ decay into a $\pair$ pair or in the decay chain $\ks \rightarrow \pi^0 + \pi^0 (\rightarrow e^- + e^+ + \gamma)$. This last channel is however suppressed by its low branching ratio $\Gamma_i$/$\Gamma \approx $1\% \cite{review-particle-physics}. A dedicated simulation performed with biased branching ratio shows that the rejection for this channel is further improved to $< 10^{-3}$ since the large emission angle of a three-body decay is significantly different from the one expected in the $\xdecay$ decay. A conservative rejection of $\sim 10^{-5}$ is reached accounting both suppression factors. As no neutral event was found using the standard criteria of $E_{S4} < 0.5$ E$_{MIP}$ in 2018 data, a number of background events of 0.006 was estimated for the calorimeter analysis \cite{visible-2018-analysis}. By adding the suppression coming from the angle using the trackers, one can conservatively estimate the background contribution from $\ks$ to be at a level of $<0.001$.

For the case of electrons, the background is expected from high energetic $\gamma$ converting in the last few layers of the WCAL. No such background was observed in a simulation of $10^7$ EOT. To estimate such contribution for $\sim10^{11}$ EOTs a data-driven method is used. A sample of 3$\cdot 10^9$ EOT was considered, roughly corresponding to $\sim$10\% of the data collected in 2018. Events in the signal region with $E_{ECAL} < 105$ GeV were selected with the requirement of at least two hits in each GEM module. Only one event with such property was found. Assuming a suppression due to the angle and minimal vertex requirement of $10^{-3}$ this would push our background down, conservatively to a level $<10^{-2}$. The analysis of the full 2018 data is compatible with this estimate: a total of three events were found with two hits in the GEM modules. For none of these events it was possible to reconstruct a physical vertex.

Table \ref{tab:dm:background} summarizes the source of background expected for this analysis. The conclusion is that the background should be under control for the full dataset accumulated during 2018 visible mode, amounting to 3$\times 10^{10}$ EOT.

\begin{table}
  \centering
  \begin{tabular}{|l|r|}
    \hline
    Background source & estimated background \\
    \hline
    $\gamma$ punchtrough from em-shower &  $<$0.01 \\
    $\pi^-$ punchtrough & $<$0.001 \\
    $K^0_S \rightarrow \pi^- + \pi^+$  &  $<$0.001 \\
    $K^0_S \rightarrow \pi^0 + \pi^0$,$\pi^0 \rightarrow \gamma + e^- + e^+$   & $<$0.001\\
    \hline
  \end{tabular}
  \caption{Background sources for NA64 visible mode tracking analysis estimated for $3 \cdot 10^{10}$ EOT}
  \label{tab:dm:background}
\end{table}

\section{Upgrade of the visible mode setup}
\label{sec:new-visible-setup}

The analysis presented in Sec.\ref{sec:new-tracking-analysis} shows that the setup used in 2018 suffers from some fundamental issues that limit its capability to probe the region of parameter space characterized by fast decaying $\dm$. Using the trackers in the current setup does not increase the sensitivity for the $\dm$ since it becomes impossible to separate the very close tracks of the $\xdecay$ decay (see Table \ref{tab:dm:efftable}). A larger distance from the decay vertex is needed to resolve the small angle of the decay. Moreover, a reconstruction of the track momenta of the $\ee$ pair is also needed to measure the invariant mass of the $\dm$ and as additional tool for background suppression. As no magnet is placed after the decay volume, it is currently impossible to perform this measurement using the trackers. One could think to exploit the transverse segmentation of the ECAL to reconstruct the two electromagnetic showers precisely (see \cite{ABBON201569} for a review of this technique). The distance between the $\ee$ pair in the ECAL plane is however just $\sim$3 mm in the current setup, and since the Molière radius for the ECAL is $\simeq$28 mm a good shower separation cannot be achieved. On top of this, larger values of the coupling $\epsilon$ suppress exponentially the detection efficiency since the short decay length decreases the probability to exit the dump. To summarize, our analysis underlines three fundamental issues to be addressed in the new setup:

\begin{itemize}
\item Increase the probability of the $\dm$ to exit the dump up to at least 20\%.
\item Increase the distance between the trackers and the decay base of the $\dm$ to allow the separation of the $\ee$ pair by at least a few mm.
\item Allow the momentum reconstruction of the $\ee$ pair in the $\xdecay$ decay with an accuracy of $\sim$1\%.
\end{itemize}

Regarding the first item, the probability to exit the dump can be increased either by raising the beam energy or by reducing the length of the WCAL. Further increase of the beam energy, unfortunately, suffers a significant drop in the beam intensity. Therefore, a primary beam energy of 150 GeV was selected as the optimal one. This means that to improve the sensitivity one needs necessarily to reduce the length of the WCAL. A new design that is able to increase the signal efficiency without impacting the background discussed in Sec.\ref{sec:new-analysis:background} is presented in Sec.\ref{sec:new-wcal}.

The two last items are to allow the reconstruction of the invariant mass of the particles in the decay volume. As the $\dm$ mass was already determined with a precision $<$1\% \cite{be8,be8-2}, the reconstruction of the invariant mass would allow an unambiguous signature to confirm the origin of the anomaly. A dipole magnet in the decay volume would allow the momentum reconstruction using the trackers. Additionally, it would also increase the distance between the $\ee$ in the ECAL, thus making the separation of the two em-showers large enough to be resolved. Still, measuring precisely the small angle between the $\ee$ in the $\xdecay$ requires particular cares. In Sec.\ref{sec:imassreco}, a technique tailored for the $\dm$ search is presented to overcome these challenges and guarantee a precision of $\sim$2\% for the $\dm$ mass. In the same section, a new setup designed to overcome all the issues discussed above is described and a detailed MC simulation to prove the capability of the setup is performed. The result is used to estimate precisely the number of EOT needed to probe completely the parameter space of the $\dm$ anomaly.

Although the three items discussed above are expected to have the largest impact, several other upgrades are in preparation for the NA64 experiment. An upgrade of the front-end electronics, trigger system, and DAQ will be performed to decrease the dead time down to 1\% to cope with the large intensity of the H4 beamline. Larger trackers with a transverse size of $\sim$250 mm will be also produced to maximize the acceptance of the $\ee$ after the magnet. Finally, a new ECAL with a larger transverse length will be produced both to increase the acceptance of the $\ee$ pair and to allow the reconstruction of their energy by separating their em-showers. The new ECAL will maintain the previous cell size (38$\times$38 mm$^2$) but will increase the number of cells in the direction parallel to the bending plane. The new design will consist of a matrix of 12$\times$6 cells, corresponding to a dimension of 438$\times$229 mm$^2$. The impact angle between the $\ee$ tracks and the ECAL was estimated to be $\lesssim$100 mrad, not significantly larger than what already measured in the current setup. Therefore, an additional hadron rejection factor of $\simeq$10 is expected using a shower profile analysis \cite{na64-prd}.

In this work, the final result is presented using the momentum reconstruction achieved with a realistic simulation of the trackers as detailed in Sec.\ref{sec:new-tracking-analysis}. The setup allows however a second method to measure the $\ee$ momentum by reconstructing their em-showers in the ECAL. In our MC we use an integrated field of 2.6 T$\cdot$m that grants a separation of $\gtrsim$8 cm ($\gtrsim$ 2 cells) between the two em-shower. This separation was estimated to be sufficient to reconstruct the original energy of the $\ee$ with $\lesssim$1\% precision. The available dipole magnets can achieve an integrated field of 3.45 T$\cdot$m corresponding to an average separation between the $\ee$ of 14 cm ($\sim$3.6 cells). This grants us additional flexibility in case a larger separation between the two showers will be needed to increase the sensitivity for the $\dm$.

\subsection{Optimization of the new WCAL calorimeter design}
\label{sec:new-wcal}

To design the new calorimeter structure, the figure of merit is the signal efficiency, which is defined mostly by the number of $\dm$ that decay outside the WCAL. This was quantified by a detailed MC simulation of the setup used to generate the energy spectrum and the decay kinematics of the $\dm$.

In the design used in previous searches, the WCAL had 34 layers in total, each of them consisting of a converter layer made of 3 mm of tungsten and an active part made of a 2 mm plastic scintillators. This sums to a total of $\sim$30X$_0$. Reducing the dimension of the WCAL would impact the radiation length used to contain the main shower and hence change the background conditions. To avoid this, the new design of the calorimeter was studied under the principle that the optimal radiation length should be approximately 30X$_0$. Three different designs were considered:

\begin{itemize}
\item An initial part of 9 layers using the original layer structure followed by an additional 25 layers of only tungsten.
\item A calorimeter consisting of 17 layers with layer-structure: 6mm tungsten + 2mm plastic scintillator.
\item A calorimeter consisting of 12 layers with a different structure: 9mm tungsten + 2mm plastic scintillator.
\end{itemize}

\begin{figure*}[tbh!]
  \centering
  \includegraphics[scale=0.5]{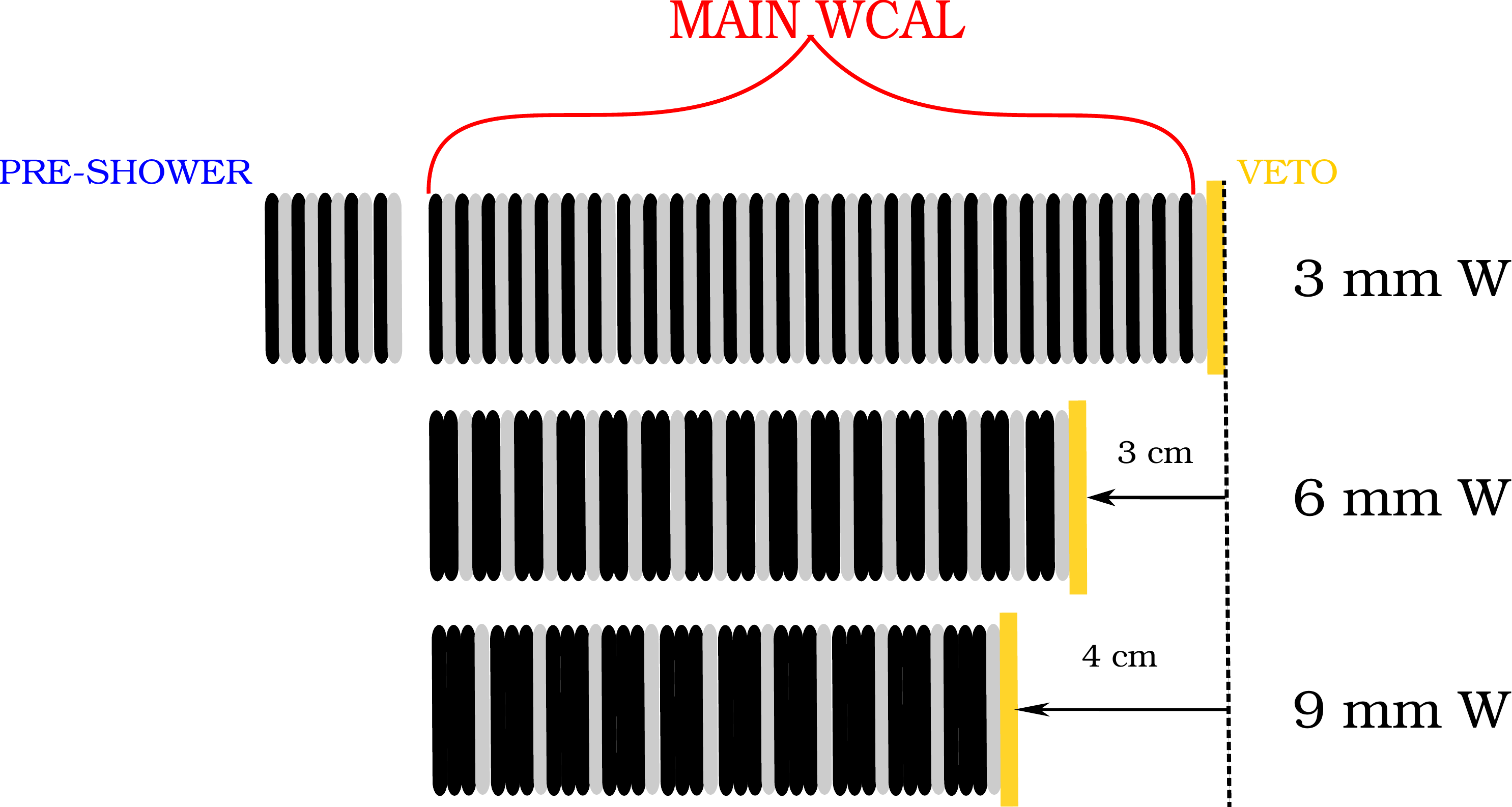}
  \caption{Possible designs of the WCAL re-arranging the available tiles of 3 mm of Tungsten (W) and 2 mm of scintillator material. All designs posses the same WCAL thickness of 30$X_0$.}
  \label{fig:wcal-design}
\end{figure*}

In all designs, the initial 5 layers forming the pre-shower part are still used for efficient hadron rejection. Despite being longer, the first design grants a good energy resolution and a good hermeticity. In the second and third case, the calorimeter is more compact but has a worse energy resolution due to the thicker converter. A sketch of the two last designs is shown in Fig.\ref{fig:wcal-design} and compared to the original one used in the previous searches.

The third design was chosen to be the most suited for our search. The loss in energy resolution has almost no impact on the signal efficiency. The reason is that the short lifetime of the $\dm$ favors the detection of the ones produced at high energy that are able to escape the dump more efficiently. These $\dm$ carry most of the initial e$^-$ energy outside of the WCAL in the calorimeter placed downstream (ECAL). Hence, the energy is reconstructed with a precision of a few \% regardless of the WCAL structure. The second and third designs are compared to the original WCAL in Table \ref{tab:wcal-length-results}.

\begin{center}
  \begin{table*}[tbh!]
    \centering
  \begin{tabular}{|l|c|c|r|}
    \hline
  WCAL structure [mm](layers) & WCAL length [mm] & $\epsilon$  & EOT to cover $\dm$ at 90\% confidence [$10^{10}$] \\
  \hline
  ECAL1:3+2(34)                & 178               & 0.001   & 17$\pm$3.4     \\ 
  ECAL1:6+2(17)                & 148               & 0.001   & 7$\pm$0.9      \\
  ECAL1:9+2(12)                & 138               & 0.001   & 6$\pm$0.7      \\
  ECAL1:3+2(34)                & 178               & 0.0012  & 85$\pm$4.7     \\
  ECAL1:6+2(17)                & 148               & 0.0012  & 24$\pm$6.9     \\  
  ECAL1:9+2(12)                & 138               & 0.0012  & 19$\pm$5      \\  
  \hline
\end{tabular}
\caption{Number EOT required to cover $\dm$ at 90\% confidence using different WCAL designs in the visible mode setup proposed for 2021. The first entry describes the structure using the convention:
  [ECAL]:[converter-depth]+[counter-depth](number-of-layers).}
\label{tab:wcal-length-results}
\end{table*}
\end{center}

\section{The $\dm$ invariant mass reconstruction technique}
\label{sec:imassreco}

The novel setup proposed for 2021 aims to further improve the background suppression and add the full invariant mass reconstruction for the decay of a very short lived particle generated at the beginning of the dump. In this section, the reconstruction technique is illustrated and the main challenges are outlined. A study based on a full MC simulation of the setup is used to demonstrate the power of the method and its capability of probing the parameter space left to justify the $\dm$ anomaly.

The remaining unconstrained parameter space for the coupling $\epsilon$  corresponds to a extremely short-lived $\dm$ with the lifetime $\tau_{\dm} \lesssim 10^{-13}$ s. If we compute the decay length of the $\dm$ we find 
\begin{eqnarray}
L_{\dm} = 28.3 ~{\rm mm}  \Bigl[\frac{E_{\dm}}{100~ {\rm GeV}}\Bigr] 
\Bigl[\frac{17~ {\rm MeV}}{m_{\dm}}\Bigr]^2 \Bigl[\frac{10^{-3}}{\epsilon}\Bigr]^2
\label{eqn:length}
\end{eqnarray}
Hence, the energy of the produced $\dm$ has to be $\gtrsim$100 GeV to have the decay length $\simeq$30 mm comparable to the dump used for the $\dm$ production in \cite{visible-2018-analysis}.
Additionally, as $E_{\dm} \gg \mee$, the minimal $\ee$ opening angle and the invariant mass are given by
\begin{equation} 
\angee^{min} \simeq  \frac{2\mee}{E_{\dm}},
\label{eqn:angle}
\end{equation}
\begin{equation}
m_{\dm} = [E_{e^+} E_{e^-}]^{1/2} \angee
\label{eqn:imass}
\end{equation}

\begin{figure*}[bth!]
  \centering
  \includegraphics[width=\textwidth]{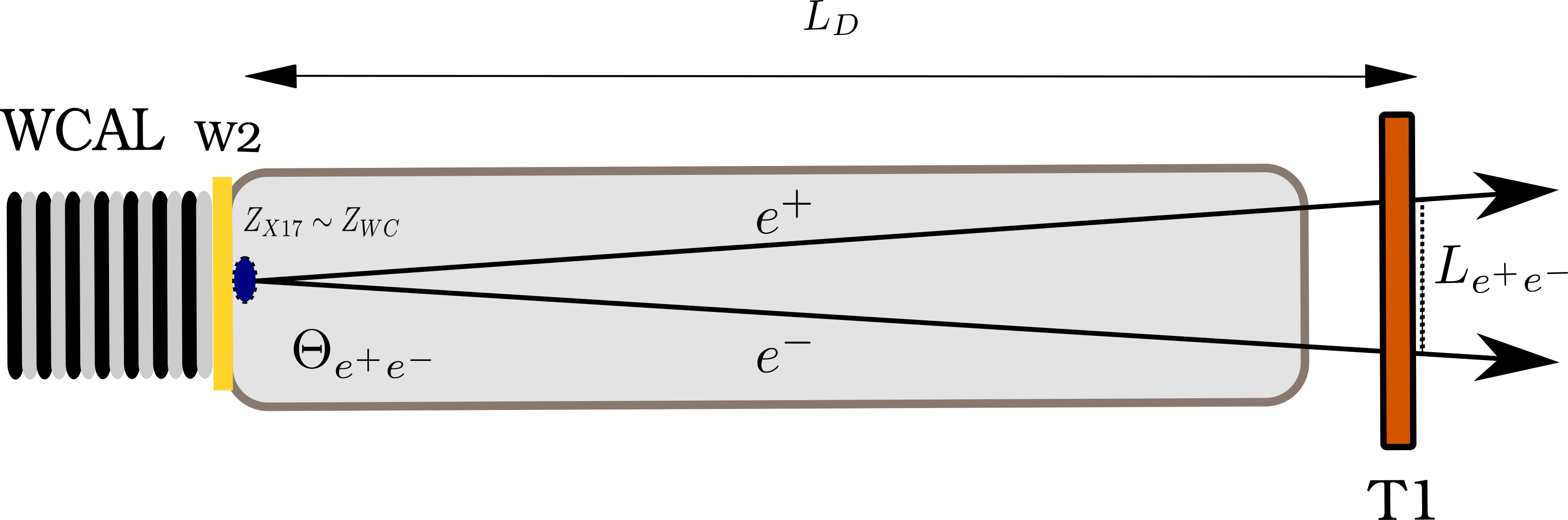}
  \caption[lol]{Sketch of the $\dm$ decay in the proposed setup along the beam axis.}
  \label{fig:imass_sketch}
\end{figure*}

For an energy $\sim$100 GeV, the average angle is $\sim$0.34 mrad, which is challenging to be measured with precision $\lesssim 10\%$. Instead, we use the short decay length to fix the vertex position of the $\xdecay$ decay to be at the end of the WCAL, and we reconstruct $\angee$ using the distance $L_{\ee}$ between the $\ee$ tracks measured by the tracker chambers placed downstream (see Fig.\ref{fig:imass_sketch}). As the $\dm$ is a short-lived particle, its decay vertex $Z_{\dm}$ is located at the vicinity of the WCAL $Z_{WC}$. This means that $Z_{\dm} \simeq Z_{WC} \ll L_D$ where $L_D=Z_{T1}-Z_{\dm}$ is the distance from the decay vertex and the first tracking detector (see Fig.\ref{fig:imass_sketch}). Since $L_D \simeq Z_{T1} - Z_{WC}$, the opening angle $\angee$ can be evaluated as 
\begin{equation}
\angee = \arctan{\frac{ L_{\ee}}{L_D}} \simeq \frac{ L_{\ee}}{L_D}
\label{angle-est}
\end{equation}
where $L_{\ee}$ is the distance of the $\ee$ pair in the T1 plane. Using error propagation, we can estimate the uncertainty on the angle:

\begin{equation}
  \sigma^2_{\angee} \simeq (\sigma_{L_{\ee}}  / L_D)^2 + (\sigma_{L_D} / L_D)^2(L_{\ee} / L_D)^2,
  \label{eqn:thetaerr}
\end{equation}

where $\sigma_{L_{\ee}}$ is the hit resolution of the tracker and $\sigma_{L_D}$ is the error of the decay base, which is the standard deviation of the distribution of the $\dm$ decays after the dump ($\simeq$ 4 cm). In our conditions, the second term is negligible due to the large distance between the trackers and the target. The formula above shows that a tube of $\sim$10 m is sufficient to reconstruct the invariant mass with a precision $\lesssim$10\%. However, this estimate is flawed by the fact that hit resolution worsens as the two hits are closer.

\begin{figure*}[tbh!]
  \centering
  \includegraphics[width=\textwidth]{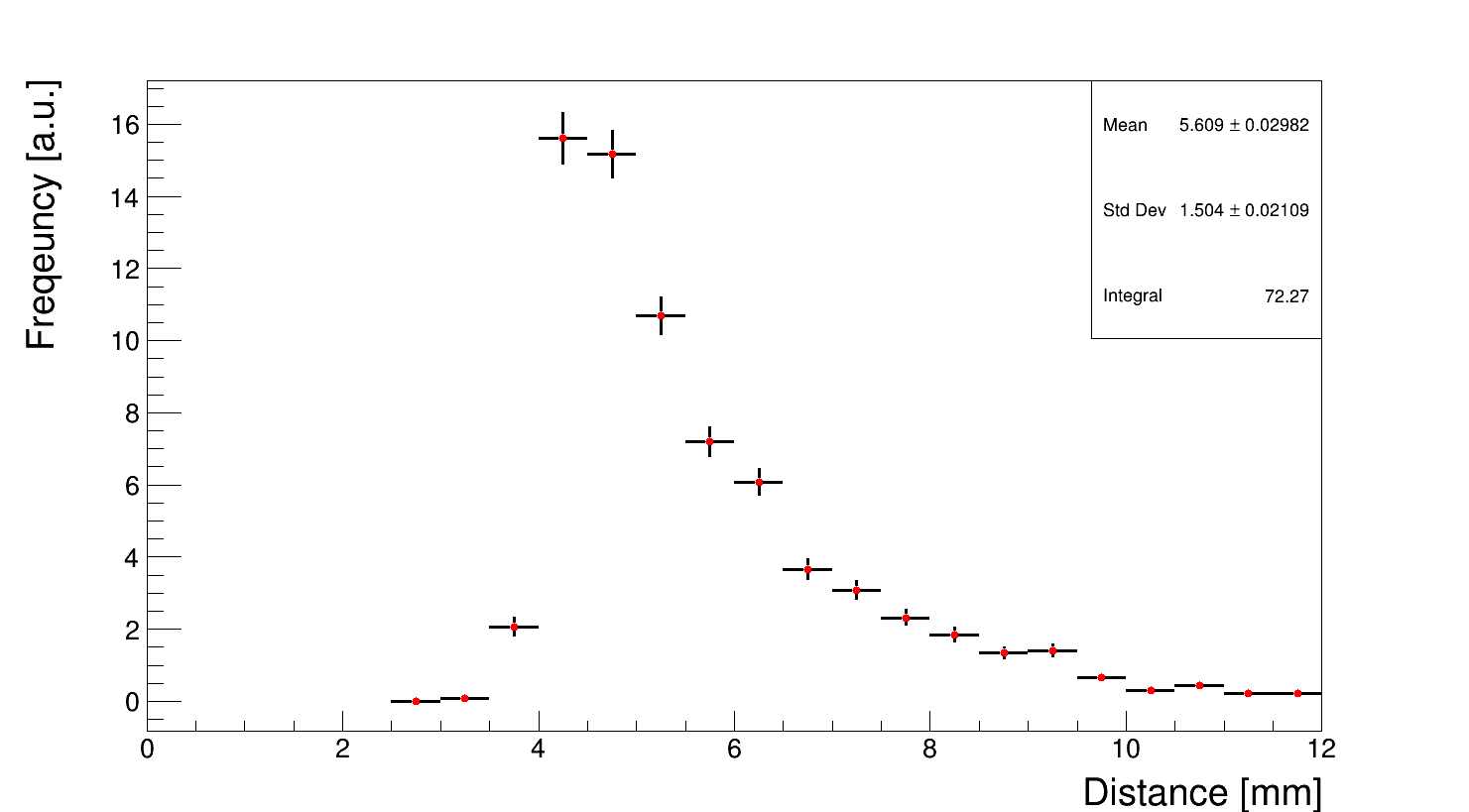}
  \caption[lol]{Distribution of the distance between $e^-$ and $e^+$ tracks from the $\dm$ decay outside the dump at a distance of 18 m from the decay vertex. The simulation was performed using a mass $m_{\dm} = 16.7$ MeV and a coupling $\epsilon = 1.4 \times 10^{-3}$ inside the proposed setup for 2021.}
  \label{fig:dm_dist1}
\end{figure*}

This problem has been studied using both fitting procedures and neural networks to reconstruct the original hit position from two overlapped clusters. The data recorded with a gas detector during past NA64 runs were used to build a set of different possible topologies. A new set to test different algorithms was then created by mixing these clusters randomly. An example of such a study, where the two clusters are separated using a global fit of two Gaussian is presented in Sec.\ref{sec:separ-hit-micr}. Both procedures agree that the hit resolution worsens to a maximum of 200 $\mu$m when the separation is lower than 1.5 mm. No significant worsening in the resolution is observed when the distance between hits exceeds $\sim$2 mm. In the proposed setup, a distance of 18 m is used between the dump and the first tracker, getting an average separation of 5.5 mm (Fig.\ref{fig:dm_dist1}). As our data-driven studies have shown, the hits should be well separated in each signal event, granting a hit resolution of 80 $\mu m$ for the $\ee$ pair.

To complete the invariant mass reconstruction one needs to know with high precision the momentum of the decay products in a signal-like event. Two independent measurements are used for this purpose. The first one is the momentum reconstruction of the two tracks after passing through a magnetic field. The second one is the measurement of the same two tracks energy in two well-separated em-showers in the ECAL downstream. A dipole magnet bends the two tracks and separate them to reconstruct their energy with a precision of 10\%$/\sqrt{\textrm{GeV}}$ in the ECAL. To achieve this purpose a separation of at least two ECAL cells ($\sim 8$ cm) is needed.

\begin{figure*}[tbh!]
  \centering
  \includegraphics[scale=0.42]{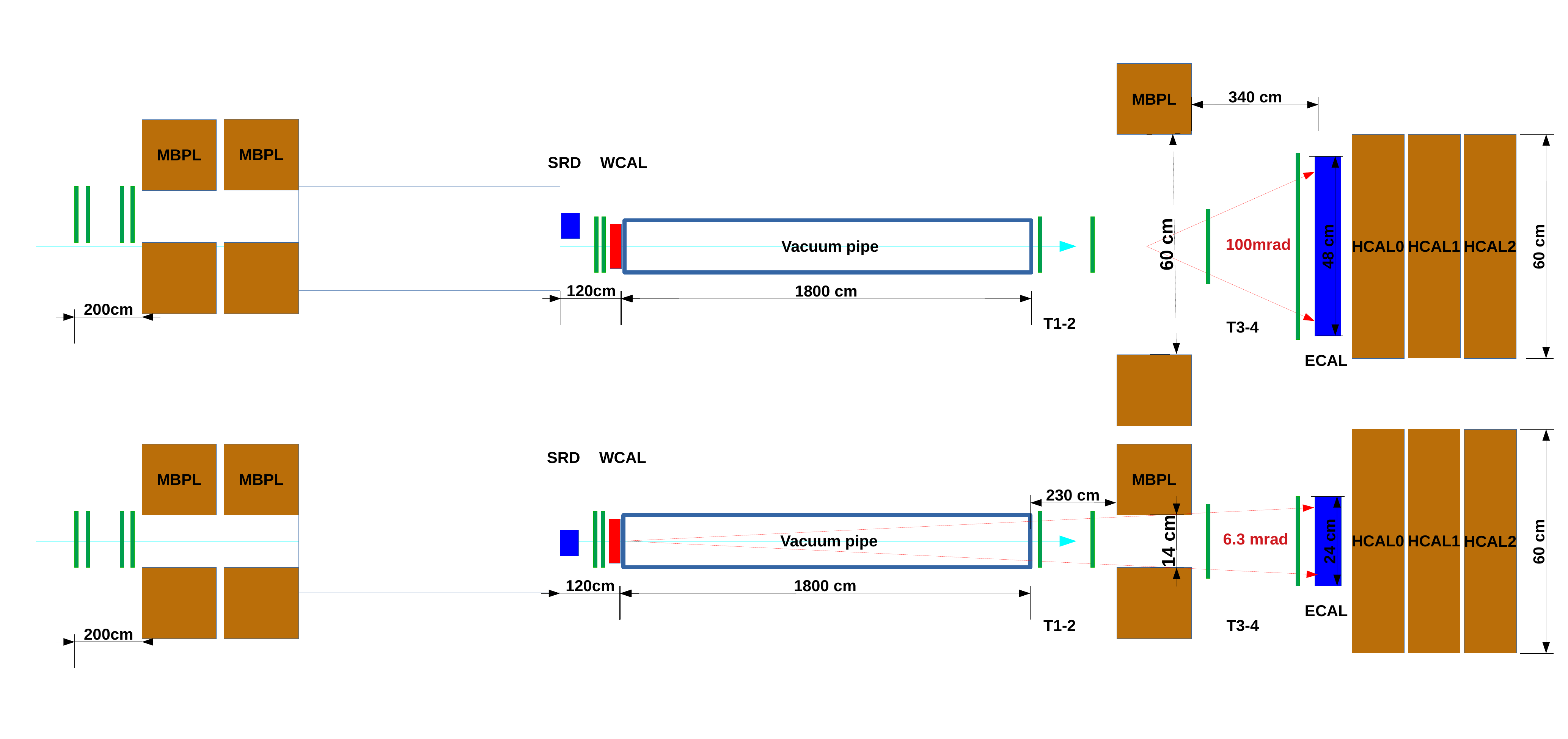}
  \caption[2021 setup]{Sketch of the setup proposed for the 2021 visible mode of NA64. Top view and side view are shown in the top and bottom pictures respectively.}
  \label{fig:setup-2021}
\end{figure*}

The setup proposed uses an 18 m vacuum tube kept at a pressure of 8$\times 10^{-4}$ mbar. Two GEM trackers \cite{gem} are placed at a distance of 0.1 m and 2.1 m respectively from the end of the tube. A magnet is placed immediately after the second GEM to separate the two tracks that are detected by a set of GEM trackers placed at 0.3 m and 1.3 m from the end of the magnet. Finally at a 3.4 m distance from the end of the magnet the ECAL is used to measure the energy of the incoming particles. A sketch of the setup can be seen in Fig.\ref{fig:setup-2021}.

\begin{figure*}[tbh!]
  \centering
  \includegraphics[scale=.8]{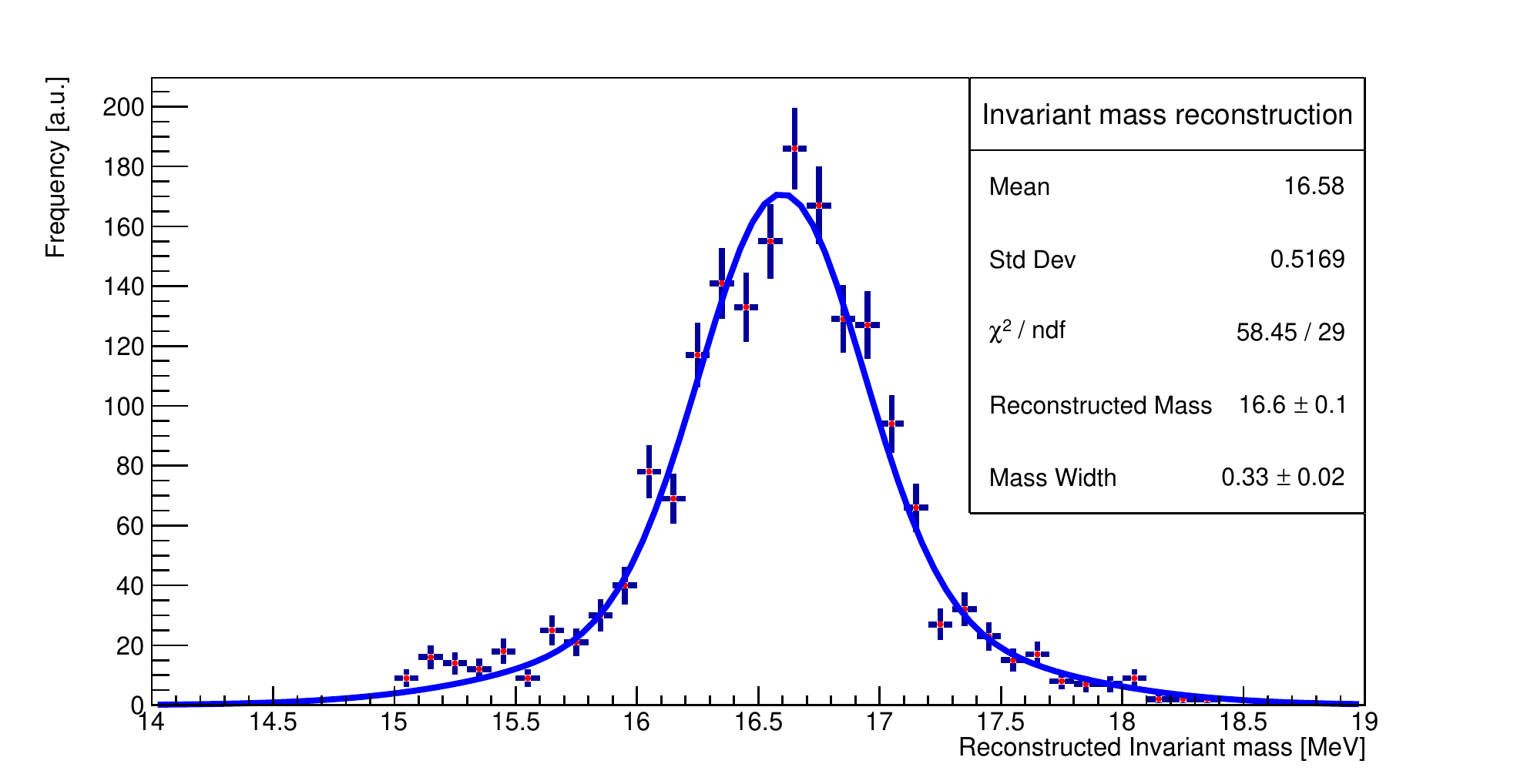}
  \caption[invmass]{Reconstructed invariant mass of $\dm$ in 2021 setup. 90\% of all events considered are reconstructed with 10\% precision. A fit performed with the sum of two Gaussian with same mean is shown as a blue line. The mass width is defined as the standard deviation of the Gaussian with largest norm. The simulation was performed using a mass of $m_{\dm} = 16.7$ MeV and $\epsilon = 1.4\times10^{-3}$.}
    \label{fig:imassreco}
  \end{figure*}

The invariant mass is reconstructed with a precision of $\sim$2\% (Fig.\ref{fig:imassreco}). The fit is performed using the sum of two Gaussian functions with a shared mean corresponding to the best estimate of the invariant mass. Furthermore, 90\% of all events are reconstructed with an error smaller than 10\%. This reconstruction was performed using a MC simulation where all detector material budget was reproduced precisely to estimate the impact of the multiple scattering. This is detailed in Sec.\ref{sec:mm-scattering}.

  \subsection{Multiple scattering effects on invariant mass reconstruction}
  \label{sec:mm-scattering}

An additional source of error is caused by the multiple scattering experienced by the $\ee$ pair produced from the $\dm$ decay. As the decay takes place immediately after the dump, the multiple scattering experienced originates from:

\begin{itemize}
\item The air pocket between the end of the WCAL and the beginning of the vacuum tube.
\item The two Mylar windows used to seal the vacuum tube.
\item Residual gas in the 18 m vacuum tube.
\item The air pocket between the tube and the trackers used to measure the distance of the two decay products.
\end{itemize}

Additionally, one has to consider that the thickness of W2 placed after the WCAL can also have an impact on the multiple scattering. This effect is however suppressed since most of the $\dm$ where the decay vertex is inside the W2 are normally removed from the analysis by the requirement of small energy deposit in this active area. The thickness of this counter was minimized to 3 mm from the 6 mm used previously. This reduces the contribution of multiple scattering and at the same time increases the $\dm$ detection efficiency since the dump length is further reduced.

The vacuum tube is placed attached to the WCAL aluminium box to minimize the air pocket down to $\sim$1 mm. Moreover a thin 175 $\mu$m Mylar window is used to seal the vacuum tube which is then kept at a pressure of 8$\times 10^{-4}$ mbar. The first detector is placed immediately attached to the vacuum tube to reduce the air interaction to a minimum. The second Micromegas tracker is placed at 2 m distance from the first one to compromise between angle and momentum resolution. All the materials were added in the MC simulation of the setup and their effects were studied in detail. The conclusion of this study is that the multiple scattering has a small impact on the precision of the reconstructed invariant mass, the degradation observed compared to a scenario where only perfect vacuum is present between the end of the WCAL and the first tracker is $\sim$0.1\%. The contributions on the invariant IMD (Invariant Mass Distribution), including limited position resolution and momentum reconstruction, are summarized in Table \ref{tab:imass-width}.

\begin{center}
  \begin{table}[bth]
    \centering
    \begin{tabular}{|l|r|}
      \hline
      Error source & Mass IMD [MeV]\\
      \hline
      Setup in vacuum & 0.11\\
      Trackers hit resolution & 0.29\\
      Vacuum window + air & 0.31\\
      Momentum reconstruction & 0.33\\
      \hline
    \end{tabular}
    \caption{Width of the invariant mass distribution after different error contributions are added cumulatively to the simulation. In the first entry, all the space in the decay volume is substituted by perfect vacuum, the only material left is the one of the trackers and the W2. In the second entry, a 80 $\mu$m hit resolution is added to the trackers. In the third entry, the vacuum is substituted by the realistic setup shown in Fig.\ref{fig:setup-2021}. Finally, the last entry add the effect of the momentum reconstruction. The invariant mass distribution with all effects considered is presented in Fig.\ref{fig:imassreco}.}
    \label{tab:imass-width}
  \end{table}
\end{center}

\begin{figure*}[tbh!]
  \centering
   \includegraphics[scale=0.75]{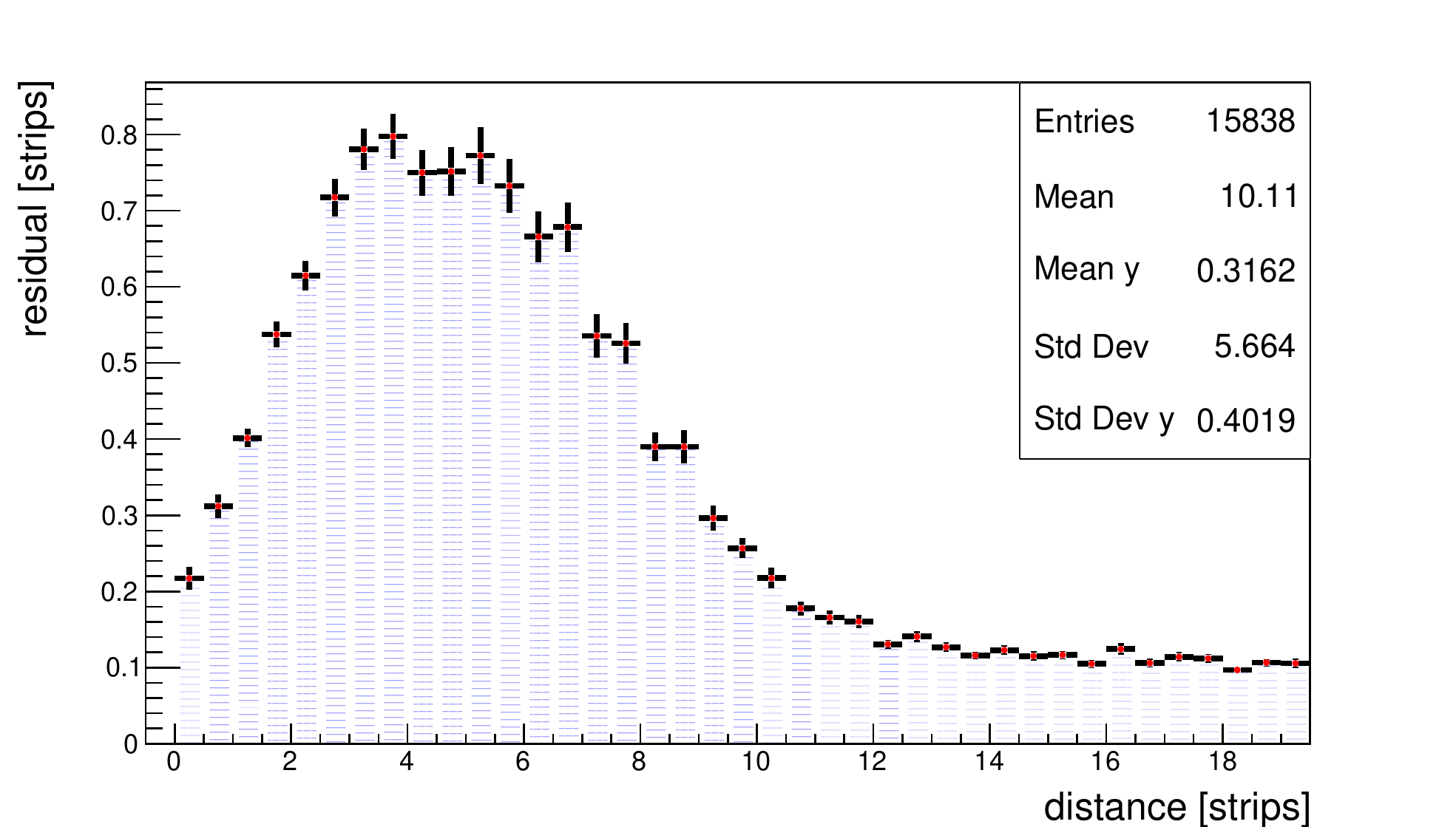}
  \caption[res-hit]{Hit resolution of two separate clusters in a same plane as function of the distance between the two. The unit are given in strips size, where a single strip has a size of 256 $\mu$m for the Micromegas used in the NA64 experiment. The hit resolution is calculated by mixing single clusters extracted from a low-intensity calibration run recorded in 2018.}
  \label{fig:res-hit}
\end{figure*}

\begin{figure*}[tbh!]
  \centering
  \includegraphics[scale=0.25]{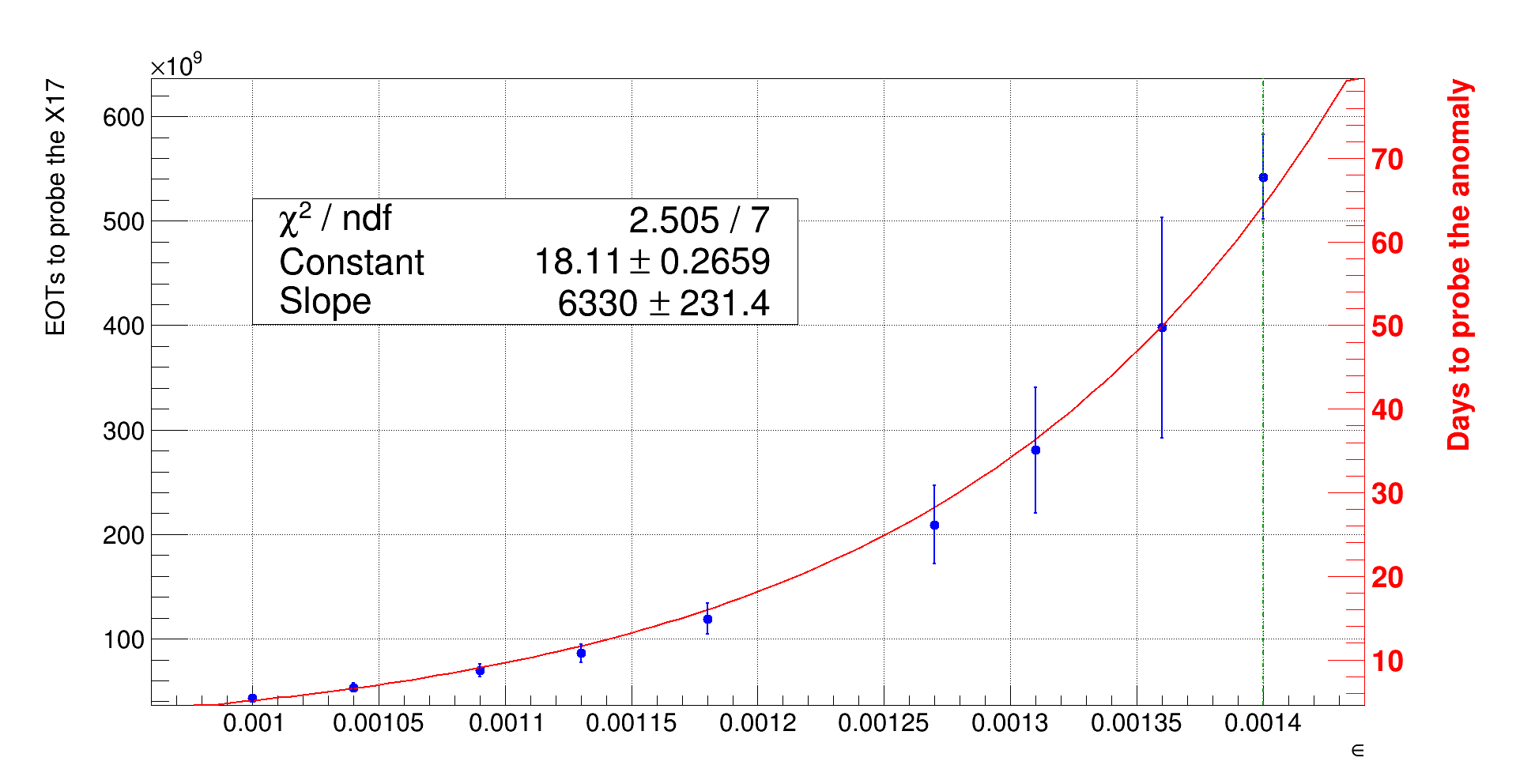}
  \includegraphics[scale=0.25]{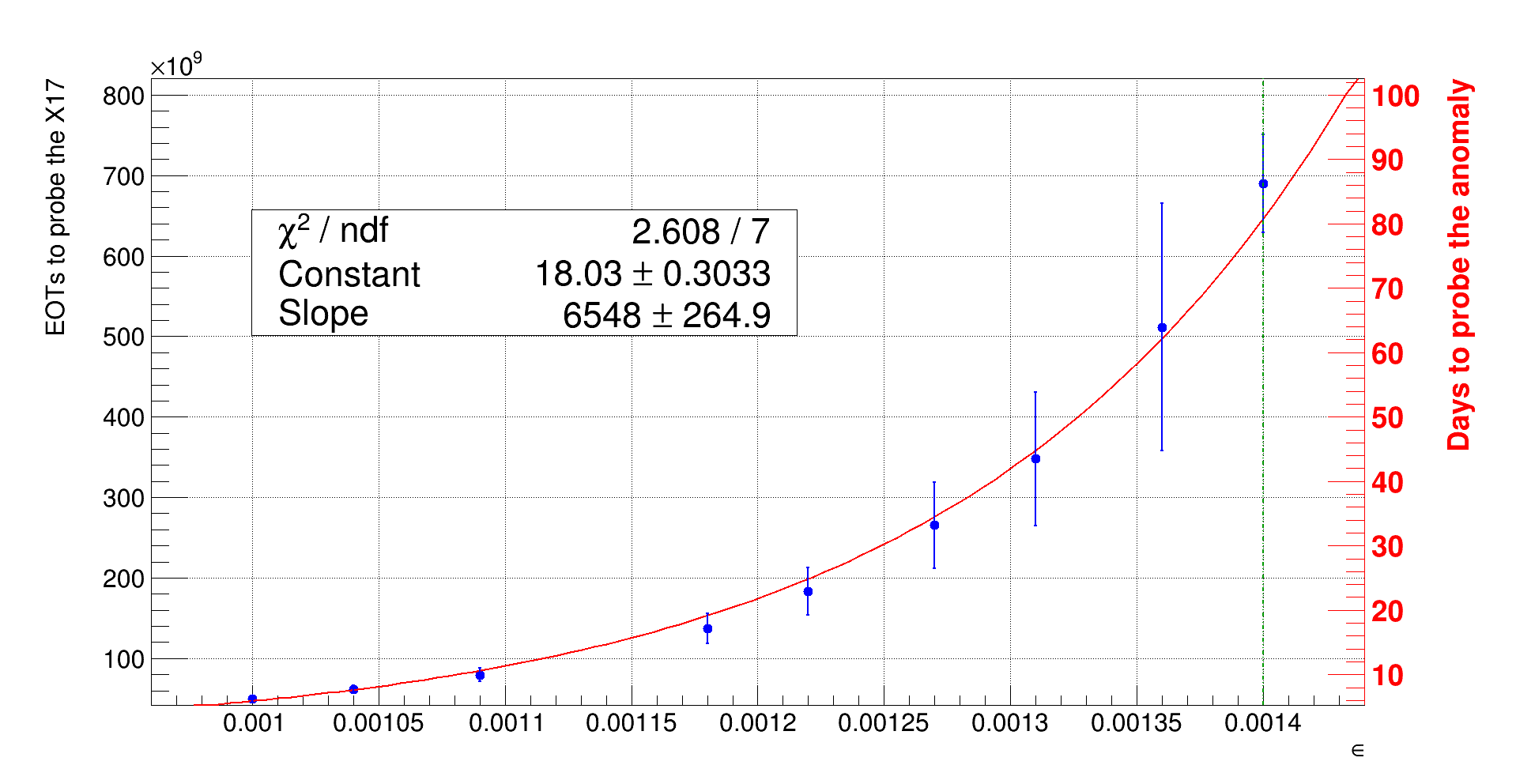}
  \caption{Number of EOTs needed to probe the $\dm$ at 90\% C.L. assuming zero background 
as function of $\epsilon$ on the left y-axis, while the number of days required to 
accumulate the correspondent number of EOTs is shown in the right y-axis and is based on 
the trigger-rate measured during the 2018 visible mode data 
taking \cite{visible-2018-analysis}. A green dashed line shows the maximum $\epsilon$ 
permitted if $\dm$ is interpreted as protophobic gauge boson \cite{feng2}. 
The detection efficiency for high $\epsilon$ is dominated by the probability of $\dm$ 
to exit the dump as it is shown by the exponential fit (red line). 
The plot is shown for the two most relevant mass scenarios suggested by the two experiments 
conducted by the ATOMKI group, i.e. 16.7 MeV (top) and 17.0 MeV (bottom) \cite{be8,be8-2}.}
  \label{fig:exclusion-x17}
\end{figure*} 

\subsection{Hit separation in gas tracking detectors}
\label{sec:separ-hit-micr}

The NA64 experiment uses gas tracking detectors to reconstruct the incoming momentum of the electrons and reconstruct tracks in the decay volume. A set of 8 XY-multiplexed Micromegas and 4 GEM modules were employed in the invisible and visible mode setup for this purpose. As introduced in Sec.\ref{sec:2018-setup}, one of the main challenges of the novel setup design will be to separate two tracks at low distance in order to reconstruct the angle of the two-body decay.

A set of clusters was extracted from the calibration data at low intensity to ensure that only single-hit clusters were present in the sample considered. The true position of these particles were saved before randomly mixing the clusters in a new set mimicking events where two particles are hitting the trackers simultaneously. After this, a double gaussian fit was used to extract the position of the two initial clusters, and the results of such procedure were compared to the known initial positions. The fit was performed using the Minuit2 minimizer implemented in the ROOT framework \cite{root}.

The results are summarized in Fig.\ref{fig:res-hit} where the hit resolution, defined as the mean difference between reconstructed hit and the true one, is shown as function of the hit separation. The results here are presented in strip size to present the problem in a general way. The part of the curve where the distance between the two clusters is between 2 and 8 strips shows a reduced hit resolution. The reason is that in this region the resulting cluster shape is significantly distorted and the fit accuracy decreases. For very close distances on the other hand, the cluster shape converges again to the one of a single gaussian, improving the fit result. In the specific situation of the NA64 experiment, Micromegas have a strip size of 256 $\mu$m, which make the two clusters separated at 9 strips ($\sim$2.3 mm). Some events with reconstructed hits exceeding a residual of 1 mm can be found for a separation smaller than 10 strips. These hits are typically caused by some abnormal cluster topology that break the gaussian assumption used by the fit. For hits with separation larger than 2 mm no such events are observed anymore. In the setup proposed in Sec.\ref{sec:new-visible-setup}, the minimum distance between the decay products is 3 mm as shown in Fig.\ref{fig:dm_dist1}. As the separation of the decay products is predicted to be much larger than the distance where the two clusters are completely separated, the $\dm$ decay products will be resolved with an efficiency $\simeq$99\%.

\subsection{Background and sensitivity}
\label{sec:background-sensitivity}

A preliminary study of background was performed in this novel setup. As discussed in Sec.\ref{sec:new-analysis:background} the main source of background is coming from the production of $\ks$ in the WCAL escaping the dump and decaying in the vacuum tube as the $\dm$. The decay products can potentially mimic the signal either in the chain K$^0_S \rightarrow \pi^0 \pi^0$ where $\ee$ pair are produced in the $\gamma$ conversion of the photon pair into an $\ee$ or in the rare decays $\pi^0 \rightarrow \gamma e^- e^+$. To estimate the impact of such background a simulation of 5$\times 10^6$ $\ks$ was performed using the energy spectrum expected from the production of this particle via electro-nuclear interactions \cite{Uzhinsky:2013hea}.
The conservative assumption of the simulation is that the $K^0_S$ is produced in an inelastic scattering in the WCAL where all the energy is deposited inside the dump without leaving a significant signature in W2. It was found that only 3\% of the events left a shower separation in the ECAL similar to the one expected from $\dm$. Less than 1\% of the events are within the acceptance of the trackers. In the majority ($>$90\%) of the surviving events the $\ks$ has an energy $<$60 GeV. This spectrum is significantly different from the one predicted for the $\dm$, where 95\% of the spectrum is above 100 GeV with a sharp peak at 150 GeV (the nominal beam energy). Finally, no event in the sample was reconstructed with an invariant mass compatible with the $\dm$, the closest one being reconstructed at 280 MeV. This is well above any $A'$ scenario in the reach of the NA64 experiment \cite{visible-2018-analysis}. As the new WCAL design conserves the hermeticity of 30$X_0$, the background coming from $\gamma$-punchtrough (see Sec.\ref{sec:new-analysis:background}) is not expected to increase in the new setup. This contribution is hard to study in detail using MC simulation, it was however demonstrated in our previous measurements \cite{visible-2018-analysis} that a longer setup adds a suppression to this background due to the large transversal spread that these particles have. A larger suppression is therefore expected due to the longer decay volume. As both neutral-punchtrough and $\ks$ are not expected to increase, one can conservatively put the background at a level of $0.01<$ (see Table \ref{tab:dm:background}).

An analysis based on the simulated data was conducted to estimate the reach of the experiment using the proposed setup. Most of the selection criteria already applied in our previous searches were used for this study. Additionally, a good separation of at least 8 cm is required between the two electromagnetic showers and the reconstructed invariant mass is selected to be within 10\% of the expected $\dm$ mass. The expected signal yield was computed after all the cuts were applied and used to calculate the 90\% C.L. for different $\dm$ scenarios. The results of the computation are presented in Fig.\ref{fig:exclusion-x17} that shows the number of EOT necessary to probe a specific $\dm$ scenario. Assuming a trigger-rate similar to the one observed during 2018 visible-mode data taking, a projection of the days needed is also shown. As expected, the EOTs required to probe a specific $\dm$ model increases exponentially with the coupling strength $\epsilon$, since the signal yield is dominated by the probability of $\dm$ to exit the dump. The conclusion is that the complete range $\epsilon < 1.4 \times 10^{-3}$ of $\dm$ parameter space proposed in \cite{feng2} can be probed in approximately 3 months of beam time by accumulating $\sim 7 \times 10^{11}$ EOTs. Models with V$\pm$A coupling mentioned in Sec.\ref{sec:introduction} on the other hand can be covered faster ($<$10 days) due to the smaller 
allowed coupling.

\section{Conclusion}
\label{sec:conclusion}

In this paper, we propose a new technique to hunt the $\dm$ boson in NA64. This setup design was optimized to probe the remaining parameter space left to justify the $\dm$ anomaly as a protophobic gauge boson \cite{feng1} which could explain the anomalies measured by the ATOMKI group \cite{be8, be8-1}. The small angle of the $\ee$ produced in the $\dm$ decay is measured after an 18 m long vacuum tube using two gas tracking detectors. The energy of the two particles is then extracted using two independent measurements of the energy deposited in an electromagnetic calorimeter and the momentum reconstructed after a bending magnet. The invariant mass can be reconstructed with a precision of 2\% in this setup. This provides an unambigous signature in case of $\dm$ detection. The background for this search was studied in detail using MC and is expected to be under control ($<0.01$). A complete study of the separation power of gas detector performed using the data collected in the previous NA64 runs shows that our current trackers can separate the decay product of $\dm$ in all scenarios considered with an efficiency close to 100\%. After considering all these contributions, a total of $\sim 7 \times 10^{11}$ EOTs was determined to be sufficient to cover the remaining parameter space of the $\dm$ anomaly at 90\% confidence level. Finally, a novel analysis of the data collected in 2018 was performed by exploiting the trackers to boost the efficiency on the $\dm$ produced at the late stage of the electromagnetic shower. In the present setup such $\dm$ have a small probability to escape the dump and therefore account to a small part($<$1\%) of the signal yield. Even though the new tracker analysis of 2018 data did not improve the sensitivity significantly, it provides an independent and complementary confirmation of our previous results \cite{visible-2018-analysis}. Moreover, it highlighted the limitations of the 2018 setup and gave a first demonstration of the tracker approach that will be used in the next generation of this experiment.

\begin{acknowledgements}

We gratefully acknowledge the support of the CERN management and staff and the technical staff of the participating  institutions for their vital contributions. This work was supported by the Helmholtz-Institut für Strahlen- und Kern-physik (HISKP),University of Bonn, the Carl Zeiss Foundation GrantNo.  0653-2.8/581/2,  and  Verbundprojekt-05A17VTA-CRESST-XENON   (Germany), Joint Institute for Nuclear Research (JINR)  (Dubna), the Ministry of Science and Higher Education (MSHE) in the frame of the Agreement No.075-15-2020-718 ID No.RFMEFI61320X0098, TPU Competitiveness Enhancement Program and RAS (Russia), ETH Zurich and SNSF Grants No. 169133, No. 186181, and No. 186158(Switzerland), and FONDECYT Grants No. 1191103, No. 190845, and No. 3170852, UTFSM PI M 18 13, Agencia National Investigastion y Desarrollo (ANID), Programa de Investigation Asociativa (PIA)AFB180002 (Chile).

\end{acknowledgements}

\bibliography{../Bibliography/bibliography}
\bibliographystyle{../Bibliography/na64-epjc}

\clearpage

\end{document}